\documentclass[aps,pre,twocolumn,superscriptaddress,preprintnumbers,amsmath,amssymb,floatfix]{revtex4-1}

\usepackage{grffile}
\usepackage{graphicx}
\usepackage{stackrel}

\newlength{\onefig}
\newlength{\twofig}
\usepackage[usenames,dvipsnames]{color}

\begin{document}

\title{Cluster glasses of ultrasoft particles}

\author{Daniele Coslovich}
\email{Email: daniele.coslovich@univ-montp2.fr}
\affiliation{Universit{\'e} Montpellier 2, Laboratoire Charles Coulomb UMR 5221, Montpellier, France}
\affiliation{CNRS, Laboratoire Charles Coulomb UMR 5221, Montpellier, France}

\author{Marco Bernabei} 
\affiliation{Donostia International Physics Center, Paseo Manuel de Lardizabal 4, E-20018 San Sebasti\'{a}n, Spain}

\author{Angel J. Moreno} 
\affiliation{Centro de F\'{\i}sica de Materiales (CSIC, UPV/EHU) and Materials Physics Center MPC, Paseo Manuel de Lardizabal 5, E-20018 San Sebasti\'{a}n, Spain}
\affiliation{Donostia International Physics Center, Paseo Manuel de Lardizabal 4, E-20018 San Sebasti\'{a}n, Spain}

\date{\today}

\begin{abstract}
We present molecular dynamics (MD) simulations results for dense fluids of ultrasoft, fully-penetrable particles. These are a binary mixture and a polydisperse system of particles interacting via the generalized exponential model, which is known to yield cluster crystal phases for the corresponding monodisperse systems. Because of the dispersity in the particle size, the systems investigated in this work do not crystallize and form disordered cluster phases. The clustering transition appears as a smooth crossover to a regime in which particles are mostly located in clusters, isolated particles being infrequent. The analysis of the internal cluster structure reveals microsegregation of the big and small particles, with a strong homo-coordination in the binary mixture. Upon further lowering the temperature below the clustering transition, the motion of the clusters' centers-of-mass slows down dramatically, giving way to a cluster glass transition. In the cluster glass, the diffusivities remain finite and display an activated temperature dependence, indicating that relaxation in the cluster glass occurs via particle hopping in a nearly arrested matrix of clusters. Finally we discuss the influence of the microscopic dynamics on the transport properties by comparing the MD results with Monte Carlo simulations. 

\end{abstract}

\pacs{Some PACS}

\maketitle

 \section{Introduction}\label{sec:intro}

A common strategy to facilitate the study of the physical properties of complex macromolecular aggregates is to coarse-grain the intramolecular degrees of
freedom~\cite{likos_effective_2001,Likos_2006}. By using standard statistical mechanical tools, it is possible to represent each macromolecule as a
single point particle and to obtain an effective pair potential accounting for the free energy of interaction between two such macromolecules. For several
macromolecular architectures, including linear chains~\cite{louis_can_2000,krakoviack_influence_2003}, rings~\cite{narros_influence_2010,Bohn_Heermann_2010},
stars~\cite{likos_star_1998,Jusufi_Likos_2009,Huissmann_Blaak_Likos_2009}, dendrimers~\cite{likos_soft_2001,gotze_tunable_2004,Huißmann_Likos_Blaak_2011} or
microgels~\cite{Gottwald_Likos_Kahl_Lowen_2004,gottwald_ionic_2005}, the so-derived effective potentials are ``ultrasoft'', i.e., the centers of mass of the macromolecules can coincide at a modest energetic cost (of order $k_\text{B}T$) without violating excluded volume interactions between monomers.

Ultrasoft particles exhibit a more complex phase behavior than that of hard ones. This originates from the interplay between entropy, which governs
the structural properties of hard-sphere solutions, and energetic contributions arising from the fully-penetrable character of the ultrasoft particles.
The topology of the phase diagram in the temperature-density plane can be classified in two classes: reentrant or monotonic behavior.
The behavior of the Fourier components of the ultrasoft bounded potential provides a necessary and sufficient condition
for observing one or the other class~\cite{Likos_Lang_Watzlawek_Lowen_2001,likos_why_2007}. Thus, if all the Fourier components are positive the crystallization lines are reentrant. 
A complex cascade of crystalline phases is found on increasing the density and these depend on the specific ultrasoft
potential~\cite{stillinger_phase_1976,lang_fluid_2000,Pàmies_Cacciuto_Frenkel_2009,Zhu_Lu_2011}. 
On the contrary, if the Fourier transform of the potential shows negative values the crystallization lines
are monotonic in the temperature-density plane. The corresponding crystalline phases of this class
are non-conventional: the ultrasoft particles form a {\it cluster crystal}~\cite{mladek_formation_2006}.
This crystal consists of clusters of particles located in the nodes of the lattice. 
Another particular feature of this phase is that the lattice constant is density independent. A direct consequence of this property
is that the cluster population is directly proportional to the density of the fluid~\cite{likos_why_2007}.

The  generalized exponential model (GEM)~\cite{mladek_formation_2006} is a well-known example of  ultrasoft bounded potential
leading to the two former scenarios, depending on the specific parameters of the
model (see Section~\ref{sec:methods}). The cluster crystal scenario of the GEM has been confirmed in a series of computational investigations~\cite{mladek_formation_2006,likos_why_2007,moreno_diffusion_2007,likos_cluster-forming_2008,camargo_dynamics_2010,coslovich_hopping_2011}.
Detailed investigations of the phase behavior have revealed an extremely complex map of cluster crystal structures~\cite{Zhang_Charbonneau_Mladek_2010}.
Some of these works~\cite{moreno_diffusion_2007,likos_cluster-forming_2008,camargo_dynamics_2010,coslovich_hopping_2011} have focused on the dynamic aspects of cluster crystals, revealing interesting properties.  
The stability of the lattice, which has a non-integer average cluster population, is maintained by incessant hopping of all the
particles between the clusters. In contrast to the usual observation in glass-forming liquids~\cite{gleim_how_1998,berthier_monte_2007}, a comparison between Newtonian, Brownian 
and Monte Carlo (MC) simulations reveals a significant role of the microscopic dynamics on the long-time dynamics~\cite{coslovich_hopping_2011}.
In particular the hopping dynamics in Brownian and Monte Carlo simulations is characterized by short-range jumps,
and the long-range, highly directional jumps found in Newtonian dynamics are strongly suppressed.
Recent non-equilibrium simulations of cluster crystals reveal novel features for their rheological response~\cite{nikoubashman_cluster_2011}. 

It is worth mentioning that potentials of the cluster-crystal class have been derived for some macromolecules~\cite{mladek_computer_2008,narros_influence_2010}.
However, though a certain degree of clustering was found in concentrate solutions of such macromolecules,
cluster crystal formation has not  been observed  yet~\cite{narros_influence_2010,lenz_monomer-resolved_2011}.
The reason is that, because of increasing many-body effects at high densities, the obtained effective potentials were no longer valid at the densities for which crystallization was predicted.
Whether one can design specific macromolecules that can form  cluster crystal phases at high concentrations is still an open issue.

As usual in colloidal systems  crystallization may be avoided in some situations, leading to  amorphous states of ultrasoft particles. 
This allows to investigate the formation of glassy states for sufficiently high densities or low temperatures.
To the best of our knowledge, all investigations on this issue have been performed in systems of ultrasoft particles showing the reentrant scenario for crystallization.
The counterpart of this phenomenon in the amorphous case is a reentrant glass transition.
This feature was predicted by the Mode Coupling Theory in a coarse-grained model of star polymers~\cite{foffi_structural_2003}, though arrested  states could not be investigated
because of crystallization. A natural way of preventing crystallization is to introduce dispersity in the particle size.
This allowed to investigate glassy states of Hertzian spheres ~\cite{berthier_increasing_2010}, confirming the presence of a reentrant glass transition.
Interestingly, it has been recently shown that Gaussian spheres can form glasses at high density even in the absence of size dispersity~\cite{ikeda_glass_2011} 
and that these have a strong mean-field character~\cite{ikeda_slow_2011}.

In this work we aim to get further insight in the dynamics of ultrasoft particles 
by investigating the glassy behavior for the class of cluster crystal-forming systems.
We present extensive computer simulations of a binary mixture and a polydisperse system of ultrasoft particles interacting through
the generalized exponential potential. We investigate structural and dynamic properties around and below the clustering transition.
The introduced size dispersity is sufficient to prevent crystallization and to produce a disordered arrangement of the clusters' centers-of-mass.
We observe the signatures of an incoming glass transition, leading to a state that we denote as ``cluster glass'', akin to the dynamically arrested states observed in colloidal systems with competing interactions~\cite{sciortino_equilibrium_2004,campbell_dynamical_2005,cardinaux_cluster_2011}.  Note, however, that the GEM potential is purely repulsive and have no minima,
i.e., for  the system investigated in this work clustering is found in the absence of attractive interactions.
A detailed analysis of the dynamics reveals a progressive arrest of the clusters' centers-of-mass on decreasing temperature, with the relaxation of the particles
taking place by hopping between the nearly arrested clusters. Finally we provide indications that the role of the microscopic dynamics (Newtonian or stochastic) on the long-time dynamics may be less important in cluster glasses than in cluster crystals.

The article is organized as follows. In Section~\ref{sec:methods} we describe the investigated
model and give simulation details. In Section~\ref{sec:results} we present the simulation results and discuss
structural and thermodynamic properties (\ref{sec:results}A), as well as dynamic properties (\ref{sec:results}B). We discuss the dependence of our results on thermal history 
and microscopic dynamics in subsection  \ref{sec:results}C.
Conclusions are give in Section~\ref{sec:conclusions}.

\section{Methods}\label{sec:methods}
We investigate the dynamics of ultrasoft fully-penetrable particles by means of
extensive computer simulations of a generalized exponential model of 
index $n$ (GEM-$n$)~\cite{mladek_formation_2006}. In this model the
interactions between the two particles are given by the bounded potential
\begin{equation}
\Phi_{ij}(r) = \epsilon_{ij} \exp[-(r_{ij}/\sigma_{ij})^n] .
\label{eq:potgem}
\end{equation}
For exponents $n > 2$ the Fourier transform of the potential has negative components, and hence the monodisperse
system can form cluster crystal phases.
In this work we focus on two different values
of the exponent $n$ in Eq.~\eqref{eq:potgem}: a binary mixture with $n=4$ and 
a polydisperse model with $n=8$.  

{\it Binary mixture.---} The system is composed of a mixture of two
species $\{1,2\}$ of particles  interacting {\it via} a GEM-4
potential. The potential is cut and
quadratically shifted at a distance $r_c=2\sigma_{\alpha\beta}$ where $\alpha,
\beta \in \{1,2\}$. The ratio between the particles' diameters is
$\sigma_{22}/\sigma_{11}=1.3$, and the cross-diameter
$\sigma_{12}=(\sigma_{11}+\sigma_{22})/2=\sigma=1$ is set as the unit of length. 
In this work we focus on the case of an equimolar mixture,
i.e., with the same number of particles for both species 1 and 2.

{\it Polydisperse model.---} The system is composed of $N$
polydisperse particles of  diameter $\sigma_{i}$. Polydispersity is
introduced by means of a flat distribution of the variable
$\sigma_i$. The distribution is centered at $\sigma=1$ and the minimum
and maximum values are $\sigma_\text{min}=0.826$ and $\sigma_\text{max}=1.164$,
respectively.  Particles interact through the pair potential in
Eq.~\eqref{eq:potgem} with $n=8$. The interaction is cut off 
at a distance $r_c=1.5\sigma_{ij}$, where
$\sigma_{ij}= (\sigma_i+\sigma_j)/2$,  $\sigma_\text{min} \le
\sigma_{ij} \le \sigma_\text{max}$, and $i, j \in \{1,..,N \}$.  In order to
discriminate between particles of different sizes, we introduce three
subpopulations of particles labelled by $\alpha=1, 2,$ and 3. If we sort the
particles by increasing value of the diameter $\sigma_i$,
 we say that the particle $i$ belongs to the species
$\alpha$ if $i \in [1+(\alpha-1) \Delta, \alpha \Delta]$, with $\Delta = N/3$. Thus
$\alpha$ increases with increasing average size of the particles. 

In both the mixture and the polydisperse system we use a common energy scale
$\epsilon_{\alpha\beta} = 1$ and particle mass $m=1$. The particles are placed
in a cubic box with periodic boundary conditions.
The static and dynamic properties of these models were investigated by
means of Molecular Dynamics (MD) and MC simulations,
performed over a wide range of temperatures $T$ and of densities $\rho = N/V$,
with $N$ the number of particles and $V$ the volume of the simulation box.
Namely we used $N = 4000$ in the binary mixture and investigated the densities
$\rho = 2.0$, 3.0, and 4.0.  For the polydisperse systems we used
$N = 4394$, 4151, 4003 and 2870 for the densities $\rho = 2.0$, 3.0, 5.0 and 7.0 respectively.

\begin{figure}[t]
\includegraphics[width=\onefig]{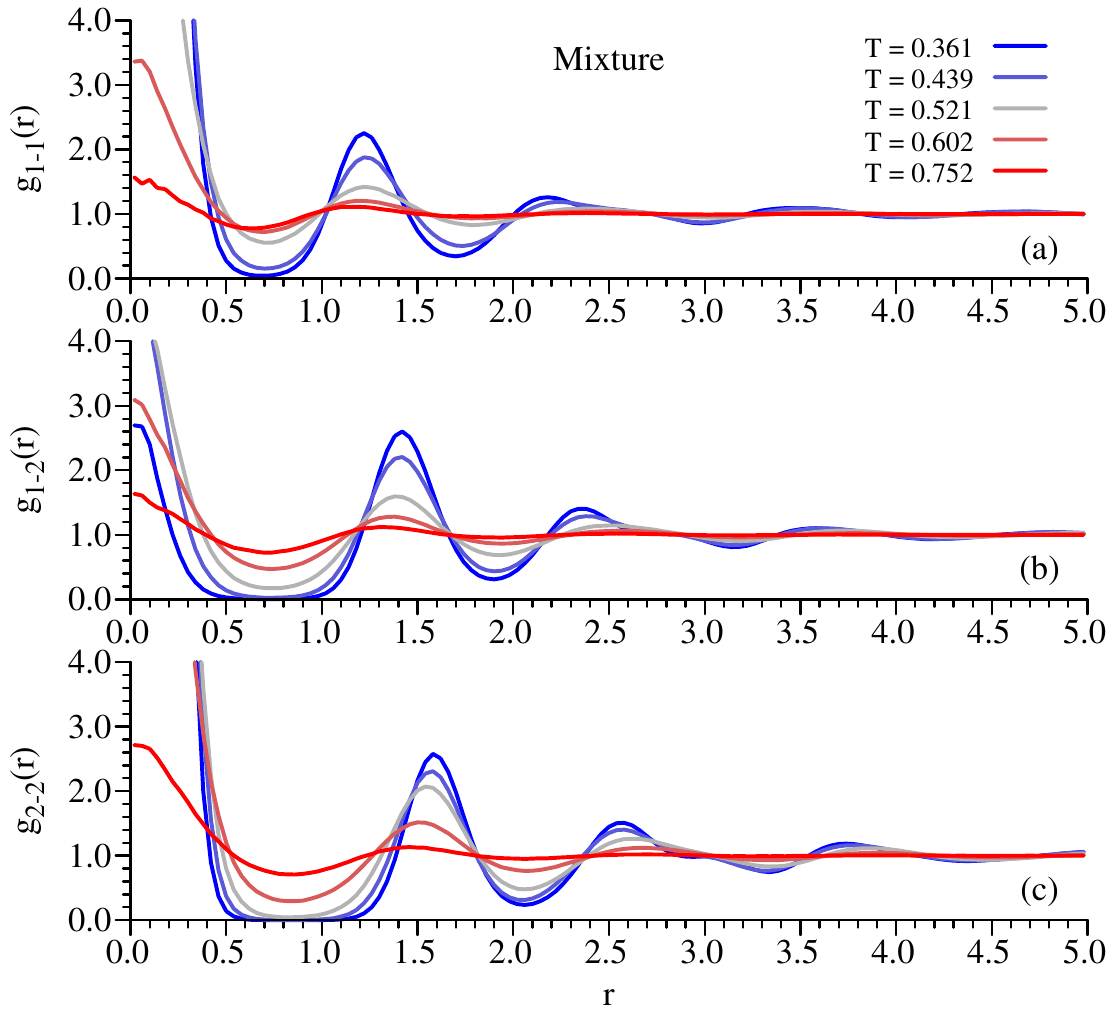}
\caption{\label{fig:gr_mix} Radial distribution functions of the
  binary mixture along the isochore $\rho=4.0$ for selected
  temperatures (see legend). (a) $g_{11}(r)$, (b) $g_{12}(r)$, and (c) $g_{22}(r)$.}
\end{figure}

Production MD runs were performed in the microcanonical  
ensemble (Newtonian MD). Newton's equations of motion were integrated by means of the
velocity Verlet algorithm~\cite{frenkel_understanding_2001}. For the polydisperse model the time step
$\delta t$ ranged from
$10^{-3}$ at high temperature to $10^{-2}$ at low temperature. For the
binary mixture the time
step $\delta t$ was 0.02 independent of temperature and
density. These values of $\delta t$ allowed to keep the degree of energy
conservation, determined from the ratio between the root mean square
deviations of the total and potential energy, to less than 2\% at all the
investigated state points.
Thermalization in the equilibration runs
was achieved by  periodic velocity rescaling in the
polydisperse model and by means of the Berendsen thermostat~\cite{allen_computer_1987} using a time constant $t_T=\delta t/0.1$
in the binary mixture. 

For comparison with the MD results we also performed
Monte Carlo (MC) simulations in the canonical ensemble. 
Propagation of the particles was implemented according to the
standard Metropolis algorithm~\cite{frenkel_understanding_2001}. The trial moves performed
during the MC simulations involved random particle displacements,
generated over a cube of side 0.1. We observed that the acceptance
ratio varied between a $80 \%$ (high $T$) and a $50 \%$ (low $T$).

\begin{figure}[t]
\includegraphics[width=\onefig]{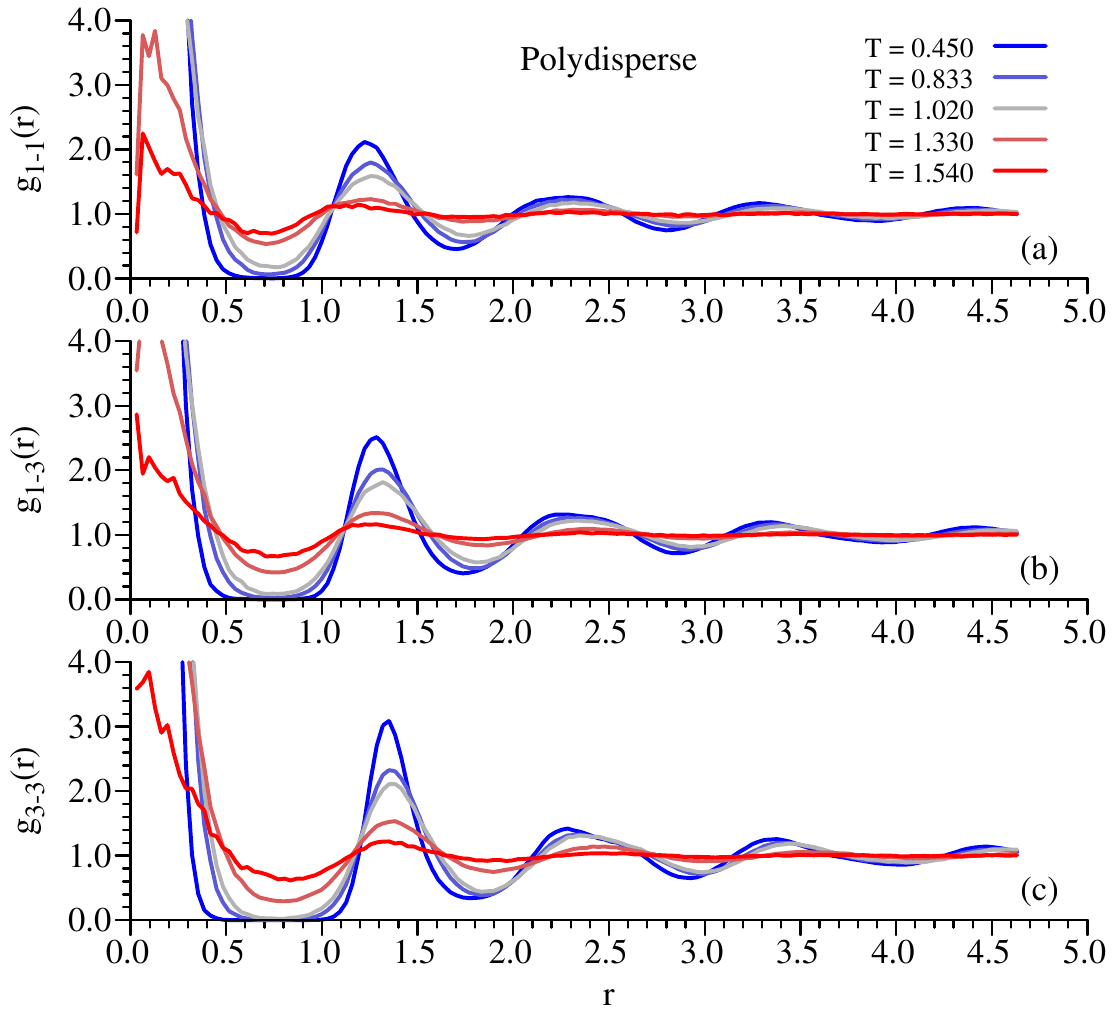}
\caption{\label{fig:gr_poly}Radial distribution functions of the
  polydisperse model along the isochore $\rho=5.0$ for selected
  temperatures (see legend). (a) $g_{11}(r)$, (b) $g_{13}(r)$, and (c)
  $g_{33}(r)$. Correlation functions involving particles of
  intermediate size ($\alpha=2$) are not shown.}
\end{figure}

To test the reliability of our results different thermalization criteria were adopted and compared. The key quantity to assess equilibration was the root mean squared displacement (RMSD)
$$ R(t) = \sqrt{\frac{1}{N} \sum_{i=1}^N|\vec{r}_i(t) - \vec{r}_i(0)|^2} \,\, , $$
evaluated at the end of the simulation. 
In the case of the binary mixture, the duration $t_\text{eq}$ of the equilibration run at any temperature was such that $R(t_\text{eq})>8$. The duration of corresponding production runs,  $t_\text{prod}$, was typically four times longer than $t_\text{eq}$, so that $R(t_\text{prod}) \approx 2 R(t_\text{eq})$. At the end of the production run, the system was cooled to a lower temperature, and the procedure was reiterated. To ensure that all the single-particle degrees of freedom, i.e., those corresponding to both small and large particles, were equilibrated, we also implemented an analogous equilibration criterion based on the \textit{partial} RMSD 
$$R_\alpha(t) = \sqrt{\frac{1}{N_\alpha} \sum_{i=1}^{N_\alpha} |\vec{r}_i(t) - \vec{r}_i(0)|^2}\,\, , $$
where the sum is restricted to particles of species $\alpha$. A small, systematic dependence on the target value $R_\alpha(t_\text{eq})$ was observed at the lowest temperatures and will be discussed in Sec.~\ref{sec:qrate}.

In the case of the polydisperse model, equilibration and production runs were such that $R(t_\text{eq})$ and $R(t_\text{prod})$ typically exceeded three interparticle diameters. In the runs at the lowest investigated temperatures, which covered about $10^8$ time steps, the target RMSD reached values of about one interparticle diameter. Even in these runs at very low temperature, however, no drift in the time dependence of the potential energy and pressure during the production runs could be observed. In addition to the same gradual cooling protocol used for the binary mixture, we also implemented an infinite-quench rate protocol, whereby the initial configuration
was always prepared by placing the particles randomly in the simulation box and then performing an infinite-rate quench to the target temperature $T$, which was subsequently equilibrated by monitoring the value of the potential energy and pressure as a function of time. We made sure that no drift was observed during the production runs. We found that the static and dynamic properties of the polydisperse model did not depend appreciably on the quenching protocol employed.

Temperature $T$, time $t$, distance $r$, wave vector $q$ and
density $\rho$ are given respectively in units of $\epsilon/k_B$ (with
$k_B$ the Boltzmann constant), $\sigma(m/\epsilon)^{1/2}$, $\sigma$,
$\sigma^{-1}$ and $\sigma^{-3}$. Unless otherwise specified, in the following the presented
results will correspond to the Newtonian MD simulations. 
The comparison between MD and MC results will be discussed at the end of Section~\ref{sec:results}.

\section{Results and discussion}\label{sec:results}

Unless specified otherwise, in the following we will present simulation results
for two selected  densities: $\rho=4.0$ for the binary mixture
and $\rho=5.0$ for the polydisperse model. As we will show below, the dynamic properties
of the models investigated here exhibit, at sufficiently low temperatures, the $\rho/T$-scaling found in the cluster
crystal phase of the monodisperse system~\cite{moreno_diffusion_2007,likos_cluster-forming_2008}. 
The scaling becomes effective for
$\rho\ge 3.0$ in the binary mixture and for $\rho\ge 5.0$ in the
polydisperse model (see below). The selected isochores are therefore
representative of the behavior in the high-density scaling regime.

\subsection{Structure and thermodynamics}

\begin{figure}[]
\includegraphics[width=\onefig]{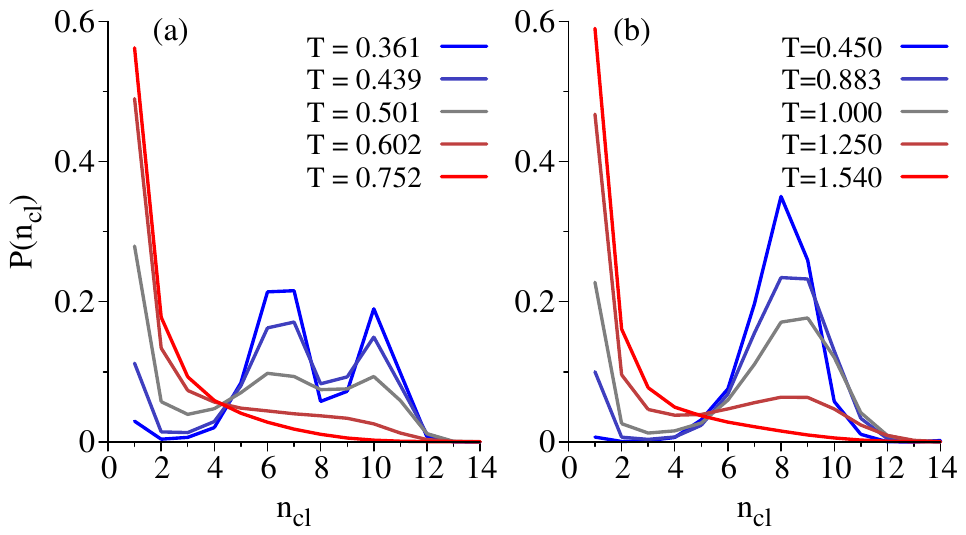}
\caption{\label{fig:cluster_distr} Distribution of cluster
  population numbers $n_{cl}$ for various temperatures (see legend) in (a) the binary
  mixture and (b) the polydisperse model.}
\end{figure}

\begin{figure}[!h]
\includegraphics[width=\onefig]{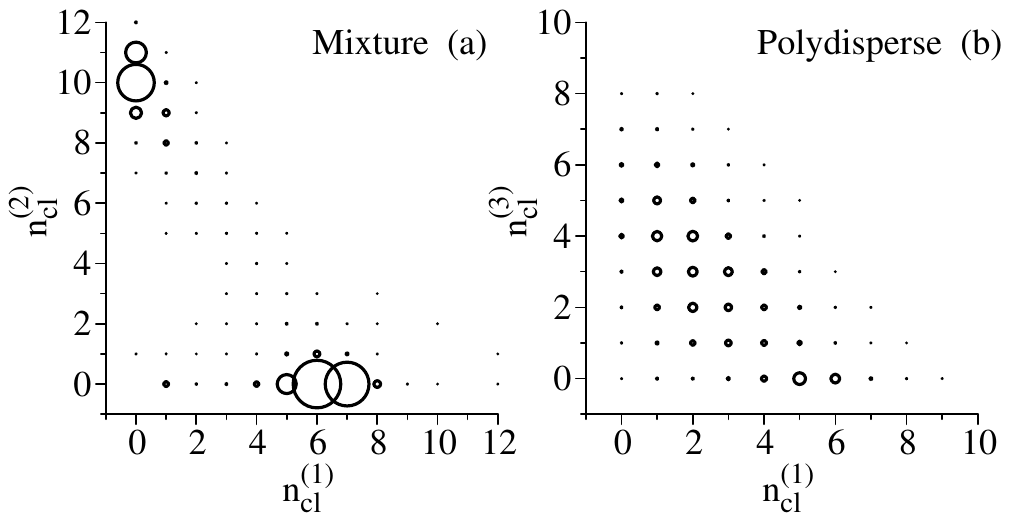}
\caption{\label{fig:cluster_partial_pops} Distribution of chemical
  compositions of the clusters (a) for the binary mixture in the $(n_{cl}^{(1)}, n_{cl}^{(2)})$
  plane  and (b)  for the polydisperse model in the $(n_{cl}^{(1)}, n_{cl}^{(3)})$ plane. The radii of the circles are proportional to the probability
  of finding clusters with a given chemical composition. The state points are (a) $T=0.35$ for the binary mixture  and (b) $T=0.45$ for the polydisperse model.}
\end{figure}

We start our discussion by analyzing the static pair correlations.
Figures~\ref{fig:gr_mix} and~\ref{fig:gr_poly} show the temperature
variation of the partial radial distribution functions
$g_{\alpha\beta}(r)$ for the binary mixture and the polydisperse
model, respectively. Both models follow a very similar structural
evolution along the selected isochores: a prominent peak located
around $r\approx 0$ builds up as the temperature decreases, indicating
an increased interpenetration of the particles. At the same time, the
first minimum in $g_{\alpha\beta}(r)$ becomes deeper, suggesting the
formation of well-defined clusters of typical maximum size $\approx
0.7$. This value can be read off from the positions of
the first minima of the radial distribution functions. The ability of the
particles to interpenetrate stems of course from the ultrasoft and bounded character of
the interaction potential, Eq.~\eqref{eq:potgem}, but formation of clusters
in purely repulsive models is a non-trivial collective phenomenon,
that arises only at sufficiently high density~\cite{likos_why_2007}. A close inspection of the $g_{\alpha\beta}(r)$ in the binary mixture shows that the peaks around $r\approx 0$ are significantly higher for like correlations. This suggests that in this latter model clusters are mostly populated
by particles of the same species---a phenomenon that can be described
as ``chemical segregation''. The data for the polydisperse model
indicate a similar tendency towards homo-coordination, although the
effect is significantly weaker than in the binary mixture.

To characterize these effects more precisely, we performed a simple
cluster analysis. A particle belongs to a given cluster if its distance to at least one of the other
particles of that cluster is smaller than a preselected cut off
$r_\text{cut}$. When using a fixed value of $r_\text{cut}$, it is
sometimes difficult to unambiguously identify the cluster to which a
particle belongs, due to the continuous flow of particles from cluster
to cluster~\cite{coslovich_hopping_2011}, in particular at high temperature. 
In practice, however, we do not
observe any major artifacts for the systems at hand. In
particular, our data indicate that merging of neighboring
clusters~\cite{coslovich_hopping_2011} is not a serious issue in our
models at sufficiently low temperature (see Fig.~\ref{fig:cluster_partial_pops}). In view of the
typical widths of the first peaks of $g_{\alpha\beta}(r)$ at low
temperature, a reasonable choice of the cut-off distance is
$r_\text{cut}\approx 0.4$ for the binary mixture and $r_\text{cut}\approx 0.35$ for the polydisperse model. Small changes of
this parameter ($\pm 30\%$) did not affect qualitatively our analysis.

\begin{figure}[t]
\includegraphics[width=\onefig]{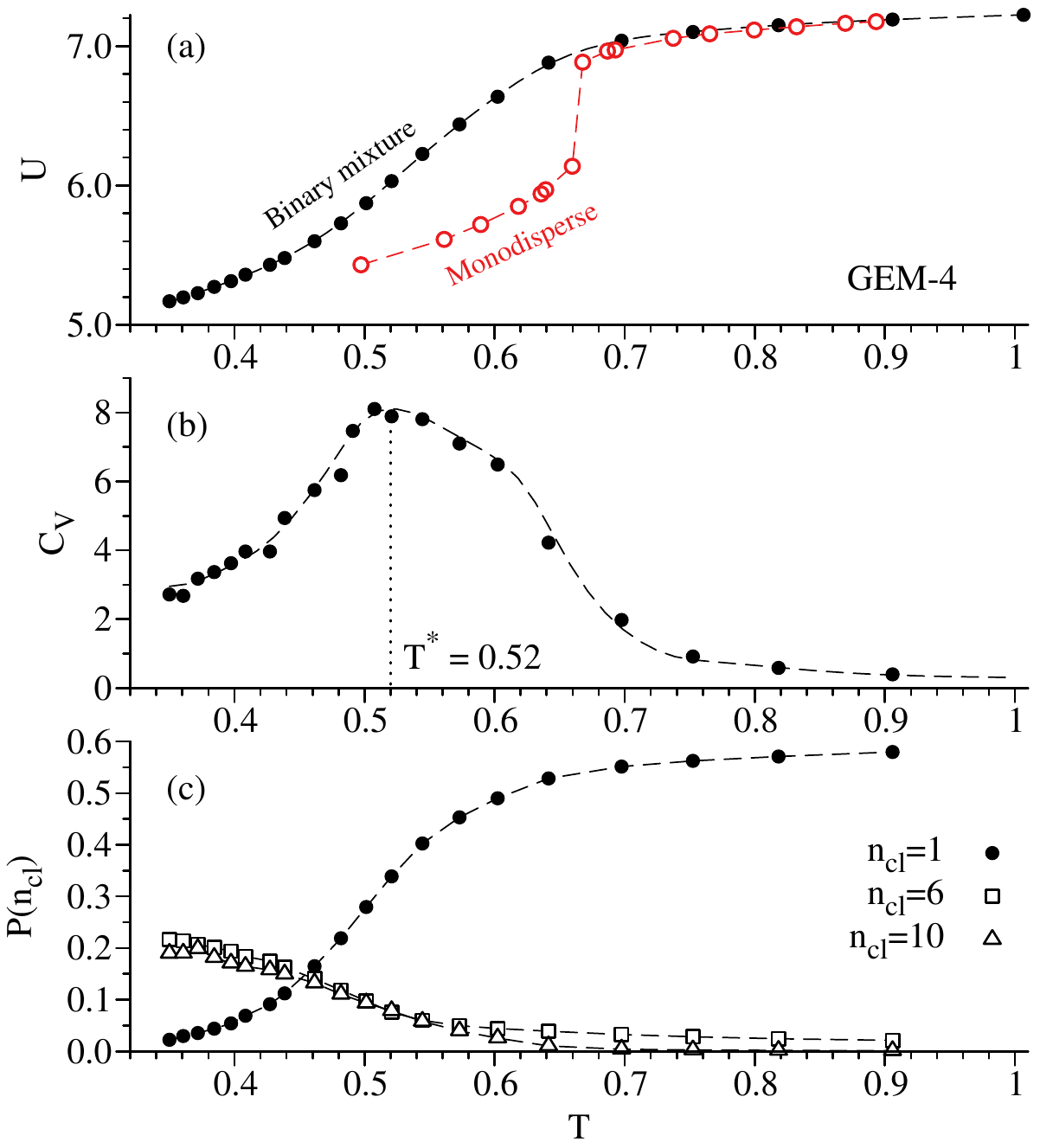}
\caption{\label{fig:cv_mix} Thermodynamic and cluster properties of
  the binary mixture (black and white symbols) as a function of $T$ : (a) total
  potential energy $U(T)$, (b) specific heat $C_V(T)$, and (c)
  fraction $P(n_\text{cl})$ of selected cluster populations $n_\text{cl}$. The vertical
  dotted line in (b) marks the position of the peak of the specific
  heat. In (a) data for the monodisperse GEM-4 model at a density $\rho=4.097$ are
  included for comparison (red symbols).}
\end{figure}

Let us first analyze the distribution $P(n_\text{cl})$ of cluster
population numbers $n_{cl}$. The temperature variation of
$P(n_\text{cl})$ is shown in Fig.~\ref{fig:cluster_distr} for the two
models. At high temperature the distribution consists of a sharp peak around
$n_\text{cl}=1$ (isolated particles), and a featureless background of more populated clusters. Thus in
the high-temperature fluid phase the system is mostly
``dissociated'', though particles may temporarily overlap due to
collisions. An analysis of the cluster lifetimes (not shown) confirms
this picture. Moreover, at such high
temperatures large values of $n_{cl}$, leading to the observed background in $P(n_\text{cl})$, may result from the fluid
structure being spatially more homogeneous, making the definition of
clusters meaningless. As the temperature is lowered, the height of the
peak at $n_\text{cl}=1$ decreases, larger clusters become more
frequent and $P(n_\text{cl})$ approaches a well-defined ultimate profile.
In the polydisperse model at $\rho=5.0$, a clear peak is
visible around $n_\text{cl} \approx 8$. In the binary mixture at $\rho=4.0$
the distribution is clearly bimodal, indicating the formation of
two distinct types of clusters with average populations of $n_\text{cl}  \approx$ 6 and 10, respectively. 
We have performed a similar analysis (not shown) at the other simulated densities.
As expected, we find that the ``preferred'' values of
$n_\text{cl}$ are density dependent, namely they tend to increase with increasing $\rho$.

\begin{figure}[t]
\includegraphics[width=\onefig]{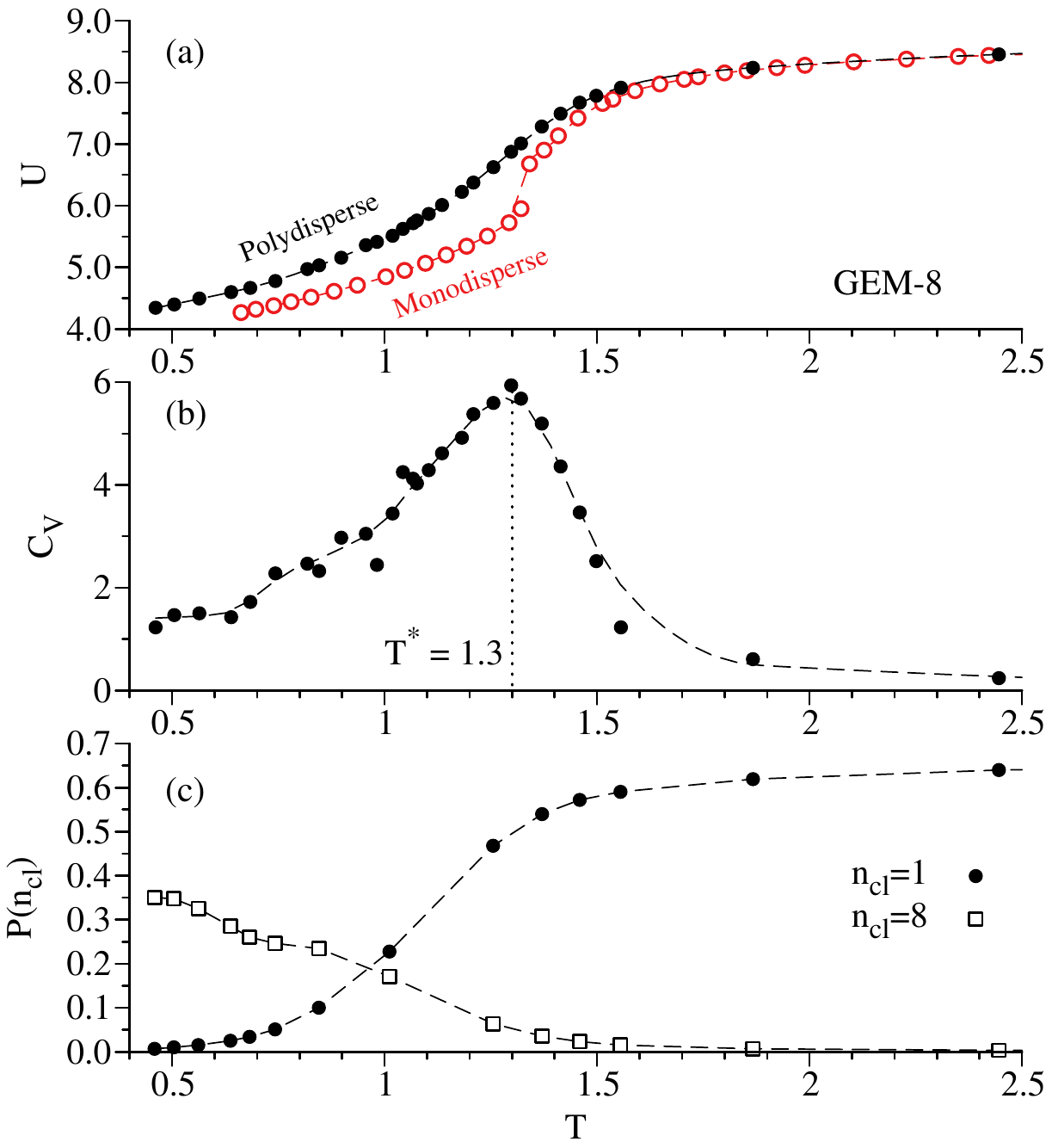}
\caption{\label{fig:cv_poly} Black and white symbols: as Fig.~\ref{fig:cv_mix}  for the polydisperse model. Red symbols in (a)
are data for the  monodisperse GEM-8 model at a density $\rho=5.0$. }
\end{figure}

\begin{figure}[t]
\includegraphics[width=0.8\onefig]{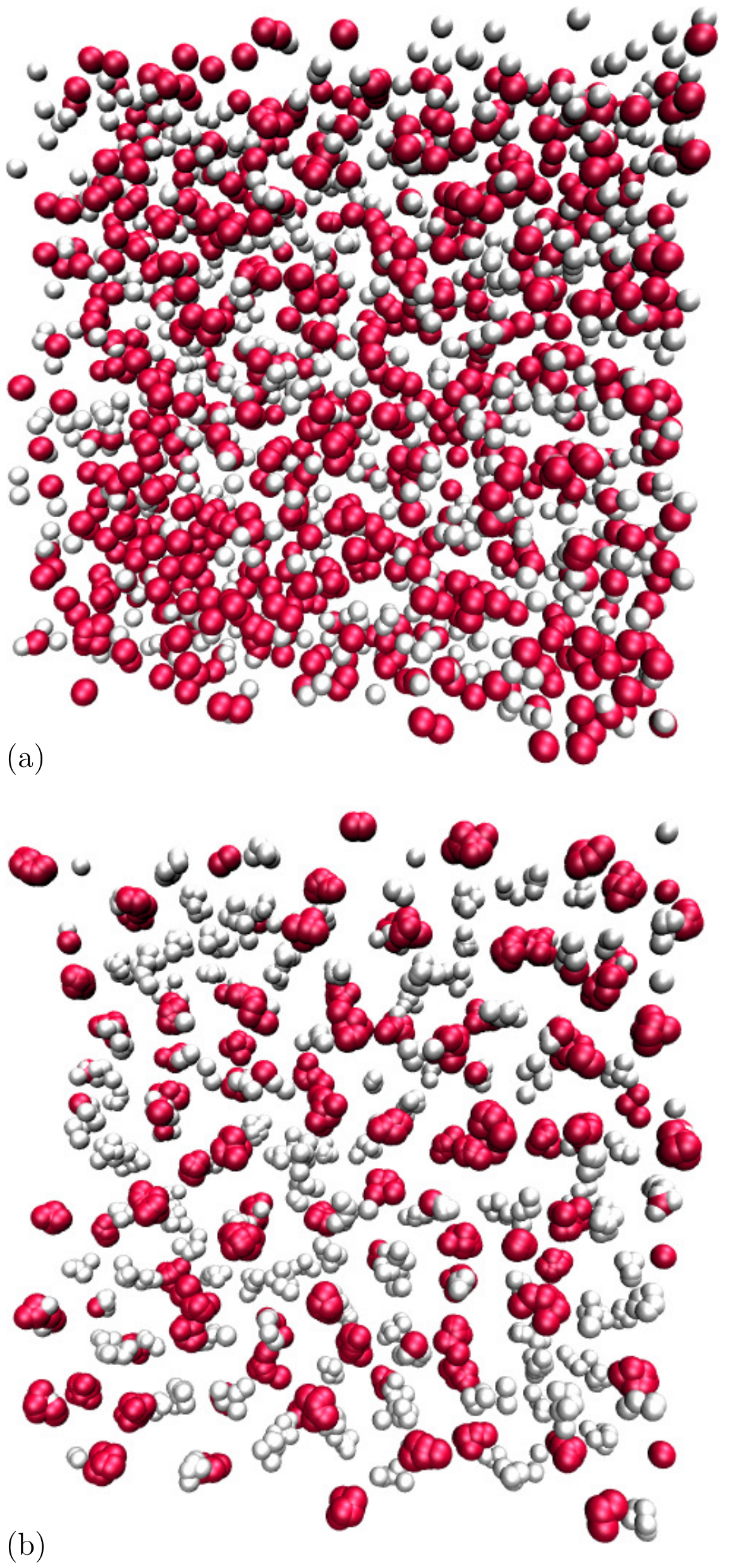}
\caption{\label{fig:snapshots_mix} Snapshots of the particles' positions, above and below the clustering temperature $T^*$ of the binary mixture: (a) $T=0.75$ and (b) $T=0.35$. Particles of species 1 and 2 are depicted as small white spheres and big red spheres, respectively. For clarity, only particles contained within a vertical slab of thickness 4 are shown.}
\end{figure}

\begin{figure}[t]
\includegraphics[width=0.8\onefig]{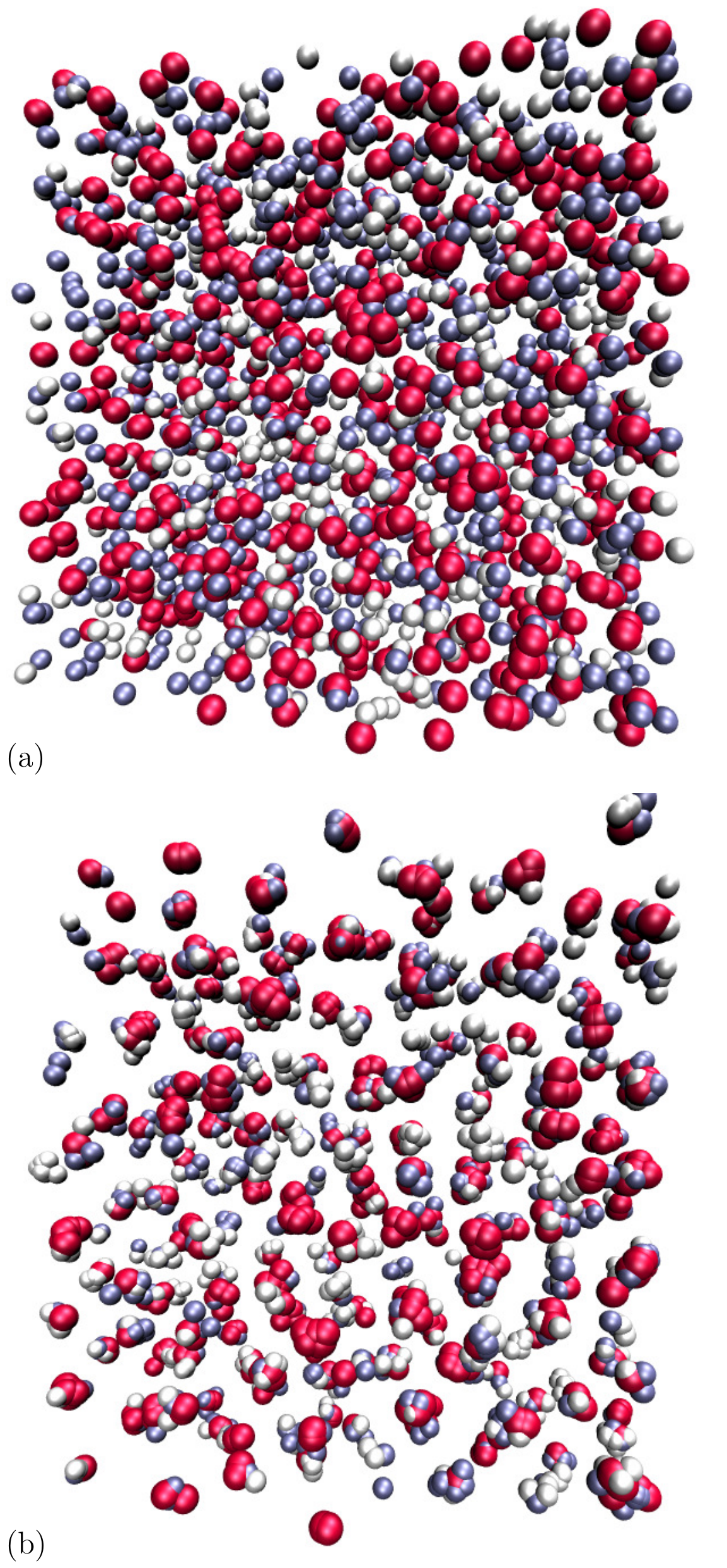}
\caption{\label{fig:snapshots_poly} Same as Fig.~\ref{fig:snapshots_mix} but for the polydisperse model: (a) $T=2.44$ and (b) $T=0.64$. Particles of species 1, 2, and 3 are depicted as small white spheres, intermediate blue spheres, and big red spheres, respectively. For clarity, only particles contained within a vertical slab of thickness 4 are shown.}
\end{figure}

A closer inspection of the chemical composition of the clusters
reveals that the double peak structure of $P(n_\text{cl})$ in the
binary mixture reflects the same chemical segregation indicated by the
radial distribution functions: ``small'' clusters ($n_\text{cl}\approx
6$) are mostly formed by small particles (species 1), whereas ``big''
clusters ($n_\text{cl}=10$) are mostly formed by big particles
(species 2). This effect is nicely illustrated in
Fig.~\ref{fig:cluster_partial_pops}. The population number of each
cluster is first decomposed as $n_{cl}= n_{cl}^{(1)}+n_{cl}^{(2)}$,
where $n_{cl}^{(1)}$ and $n_{cl}^{(2)}$ indicate the number of
particles of species 1 and 2 composing the cluster,
respectively. Then, the two-dimensional map of the distribution of
clusters with ``chemical composition'' $(n_{cl}^{(1)}, n_{cl}^{(2)})$
is constructed. The distributions shown in the figure correspond to
thermodynamic states located in the temperature regime where strong
clustering is observed. For the polydisperse model, the analysis is
performed using $(n_{cl}^{(1)}, n_{cl}^{(3)})$ and averaging over all
possible values of $n_{cl}^{(2)}$. In the binary mixture, the tendency
towards chemical segregation is rather evident. It is worth mentioning that 
at these low temperatures ``merged clusters'' are extremely rare. Cluster merging in the binary mixture would lead to the appearance of chemical compositions such as $(n_{cl}^{(1)}, n_{cl}^{(2)}) = (12,0)$ or $(0,20)$ (i.e., union of preferred clusters composed of small and big particles, respectively), or $(6,10)$ (i.e., union of two different preferred clusters). However, the distribution shown in Fig.~\ref{fig:cluster_partial_pops}a has no contributions around $(n_{cl}^{(1)}, n_{cl}^{(2)}) = (6,10)$ and (0,20). Only a few instances of (12,0) and (12,1) clusters are visible, the latter arising from two (6,0)-clusters merged by one big particle, but their fraction is negligible ($<10^{-5}$).
Thus, we conclude that the binary
mixture of GEM-4 particles self-assemble at low temperatures into a
 \textit{binary mixture of clusters}. The results for the polydisperse 
model indicate an anti-correlation between subpopulations of small and
large particles, but the effect of chemical segregation is much weaker than in the binary mixture
(see Fig.~\ref{fig:cluster_partial_pops}b). 
Thus, the clusters in the polydisperse
model remain intrinsically polydisperse in character.

\begin{figure}[t]
\includegraphics[width=\onefig]{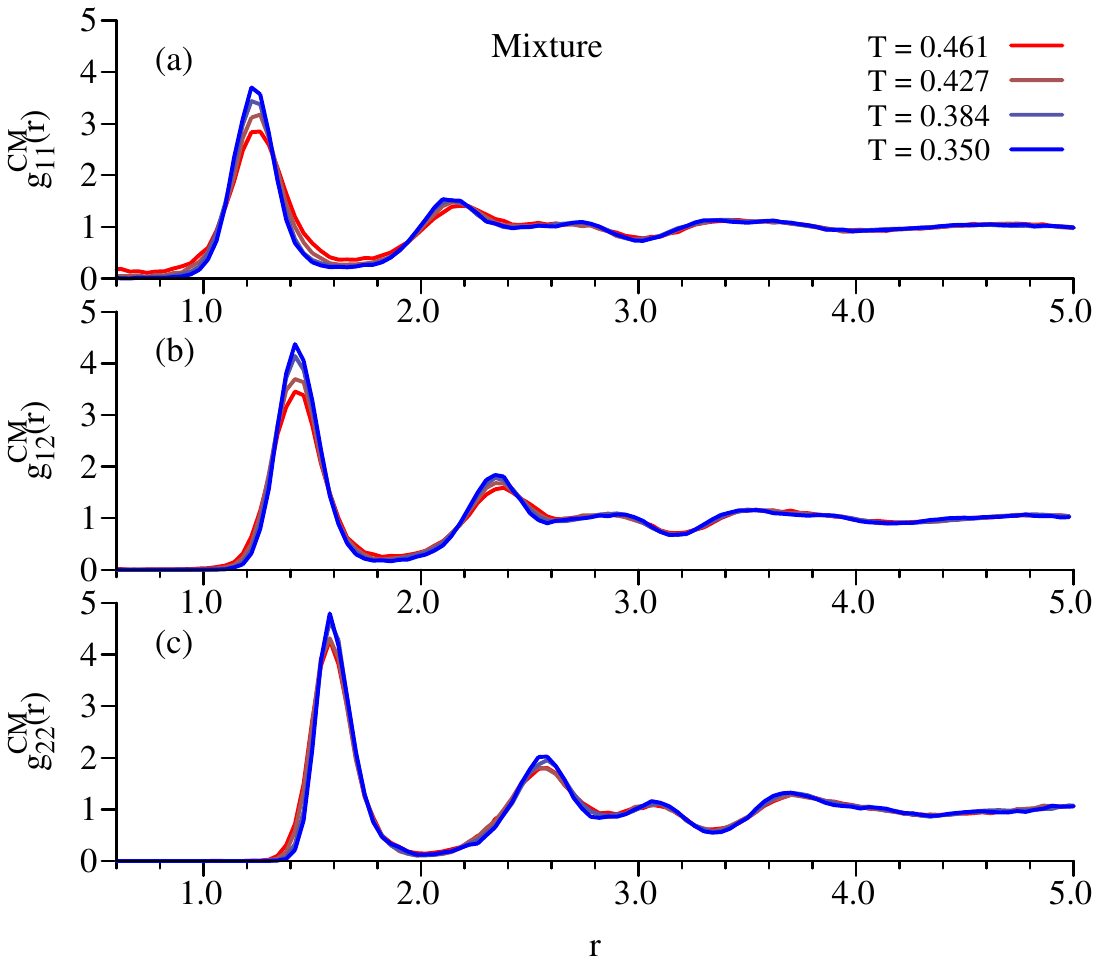}
\caption{\label{fig:gr_cluster_mix} Radial distribution function
  of the clusters' centers of mass in the binary mixture at
  some selected temperatures (see legend): (a) $g_{11}^\text{CM}(r)$, (b) $g_{12}^\text{CM}(r)$,
  and (c) $g_{22}^\text{CM}(r)$.}
\end{figure}

In the following, the crossover from the dissociated fluid phase at
high temperature to the cluster-dominated regime at low temperature
will be identified, without too much rigor, as a ``clustering
transition''. Though we have no evidence of the possible thermodynamic
nature of this phenomenon, inspection of the excess specific heat
$C_V(T)=\frac{1}{N}\frac{dU}{dT}$ indicates a clear operational
definition of the transition. 
The total potential energy per particle, $U(T)$,  and
$C_V(T)$ are shown in Figs.~\ref{fig:cv_mix} and
\ref{fig:cv_poly} for the binary mixture and polydisperse system, respectively.
Both systems display a smooth decay in $U(T)$ on decreasing $T$, and a broad peak in $C_V(T)$.
The maximum of this peak, at temperature $T^*$,  marks the clustering transition. $T^*$ takes value 0.52 and 1.3 for the binary mixture and polydisperse model, respectively.
An analogous definition of the clustering transition has been used in the
study of the microphase separation of a model with short-range attraction
and long-range repulsion~\cite{imperio_microphase_2006,imperio_microphase_2007}.
For comparison, we have included in Figs.~\ref{fig:cv_mix}-a and
\ref{fig:cv_poly}-a the respective results for the monodisperse systems (GEM-4 and GEM-8, respectively). These systems were prepared in their equilibrium cluster crystal phases and subsequently heated up to high temperatures. The density was $\rho=5.0$ and $\rho=4.097$ for the GEM-8 and GEM-4 models, respectively. The latter value of the density was chosen so as to match precisely the effective packing of the binary mixture within an ``effective one-component'' description~\cite{hansen_theory_1986}. This adjustment was necessary to achieve the expected full collapse of the potential energies of the two GEM-4 models at high temperatures.
The abrupt change in the potential energy of the monodisperse systems indicates the transition from the fluid to the cluster crystal phase, which is a true thermodynamic transition.
In the following we will denote the corresponding melting temperature as $T_m$.
Interestingly, $T^*$ of the polydisperse system is very close to the corresponding $T_m$ of the monodisperse model, whereas the difference is more pronounced for the binary mixture.

To correlate the variation of the former thermodynamic quantities to the
formation of clusters in the binary mixture and polydisperse system, we show in Figs.~\ref{fig:cv_mix}-c and ~\ref{fig:cv_poly}-c the $T$-dependence of the fraction $P(n_{cl})$ of clusters with selected
values of $n_{cl}$. The fraction of isolated particles ($n_{cl}=1$)
decreases  rapidly around  $T^*$. Concomitantly, the
fractions of the preferred clusters increase, but in a rather smooth
fashion. Therefore, although a clear signature of the
clustering transition can be found in the thermodynamic properties,
one should bear in mind that it may well represent a crossover and not
a sharp, thermodynamic transition.
Around $T^*$, clusters and
isolated particles are indeed in continuous and dynamic exchange---a
sort of ``chemical equilibrium'' picture.
Figs.~\ref{fig:snapshots_mix} and~\ref{fig:snapshots_poly} show typical snapshots of the binary mixture and polydisperse
system above and below the clustering transition $T^*$. The densities  ($\rho = 4.0$ and 5.0 for the mixture
and polydisperse system, respectively) are the same for which structural and thermodynamic observables
have been presented in previous figures. The snapshots confirm visually the general structural features
discussed above.

\begin{figure}[t]
\includegraphics[width=\onefig]{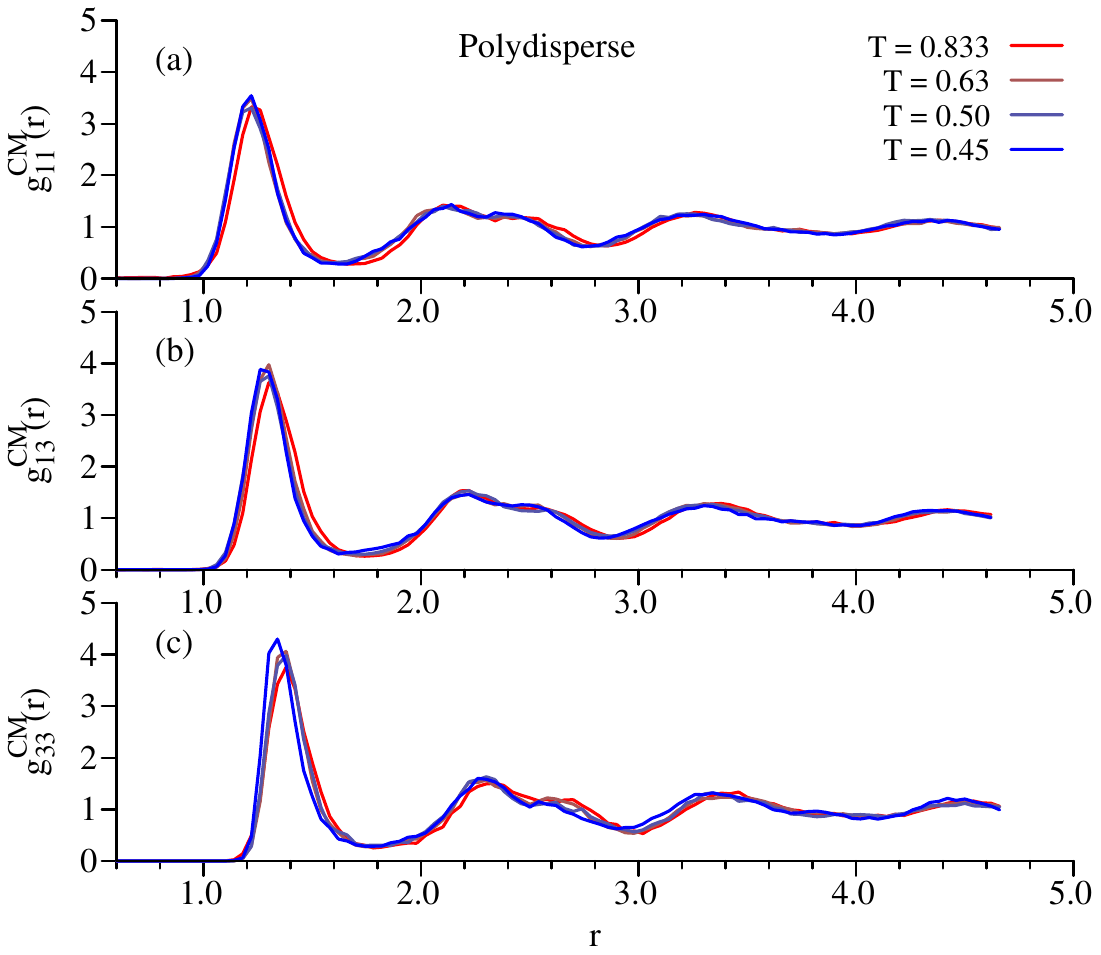}
\caption{\label{fig:gr_cluster_poly} As Fig.~\ref{fig:gr_cluster_mix}  for the
  polydisperse system: (a)  $g_{11}^\text{CM}(r)$, (b) $g_{13}^\text{CM}(r)$,
  and (c) $g_{33}^\text{CM}(r)$.}
\end{figure}

\begin{figure}[t]
\includegraphics[width=\onefig]{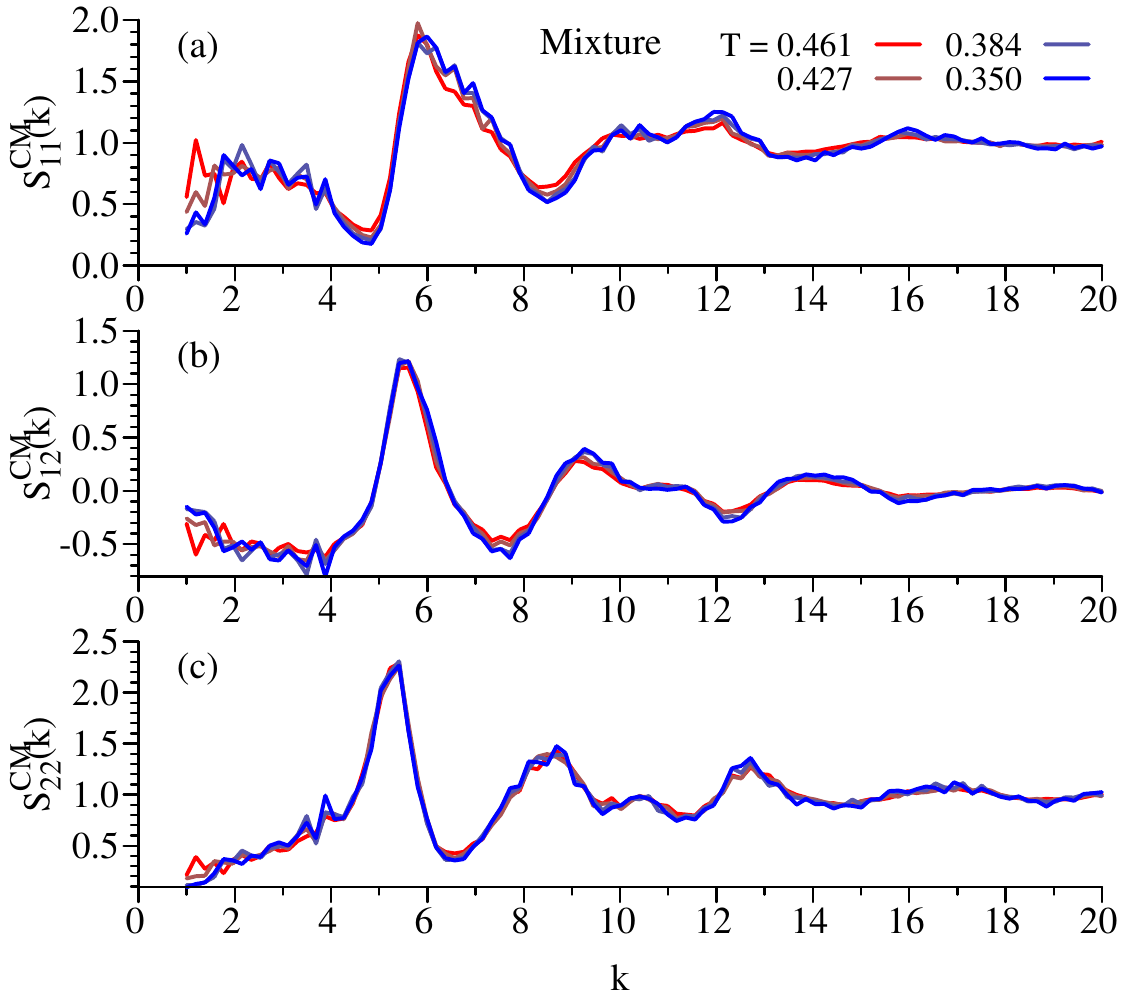}
\caption{\label{fig:Sk_cluster_mix} Static structure factors
  of the clusters' centers of mass in the binary mixture at
  some selected temperatures (see legend): (a) $S_{11}^\text{CM}(k)$, (b) $S_{12}^\text{CM}(k)$,
  and (c) $S_{22}^\text{CM}(k)$.}
\end{figure}

\begin{figure}[t]
\includegraphics[width=\onefig]{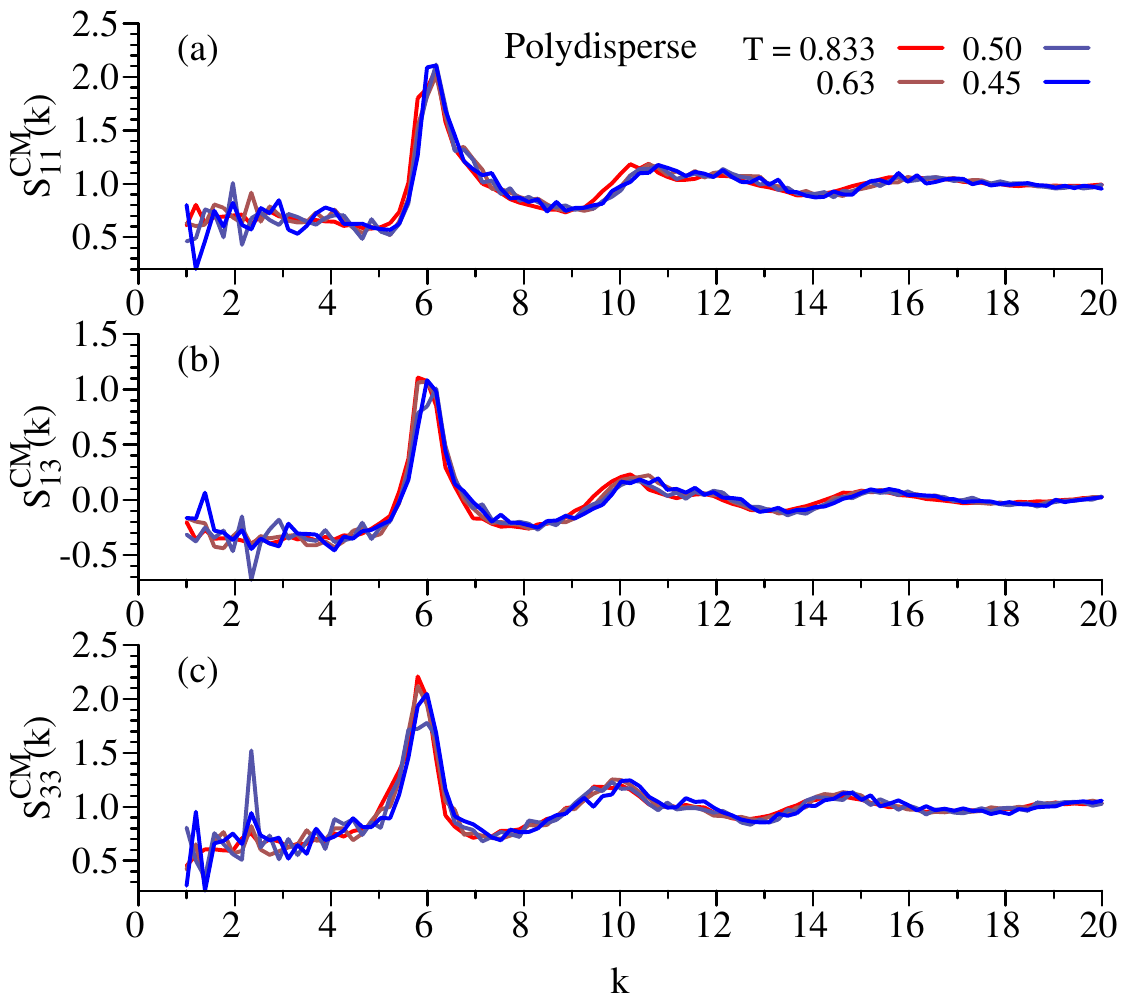}
\caption{\label{fig:Sk_cluster_poly} Static structure factors
  of the clusters' centers of mass in the polydisperse system at
  some selected temperatures (see legend): (a) $S_{11}^\text{CM}(k)$, (b) $S_{13}^\text{CM}(k)$,
  and (c) $S_{33}^\text{CM}(k)$.}
\end{figure}

As temperature decreases well below the clustering temperature $T^*$, the fraction of isolated
particles becomes negligible ($\alt 10\%$) and the system enters in a
``cluster phase'', i.e., a regime where almost all particles form
tightly bounded clusters and only rare jumps allow particles to be
transferred from one cluster to another. What is the structure of the
clusters of such a cluster phase? To address this point, we analyze
the radial distribution functions $g_{\alpha\beta}^\text{CM}(r)$ of
the clusters' centers of mass (CM). The index $\alpha$ indicates here the
``chemical composition'' of the clusters. Namely, $\alpha$-clusters are 
defined as those composed by a majority of particles
of the species $\alpha$. Isolated particles are excluded from this analysis. In Figs.~\ref{fig:gr_cluster_mix} and
\ref{fig:gr_cluster_poly} we show, for temperatures below the clustering
transition, the functions  $g_{\alpha\beta}^\text{CM}(r)$ of the binary
mixture and polydisperse model, respectively. We find that the cluster structure changes only slightly upon
cooling, without any major transformation. In particular,
the cluster structure of the polydisperse model is rather insensitive to
temperature variation and clearly amorphous. The temperature dependence of the radial distribution
functions of the binary mixture is somewhat stronger than that of the polydisperse model. Moreover these functions
exhibit a more marked splitting of the second
peak, especially for 2-2 correlations. The typical distances between neighboring clusters can be read
off from the location of the first peaks and reflect the different
``chemical'' composition: clusters composed of big particles tend to
have larger distances to their first neighbors. 
In Figs.~\ref{fig:Sk_cluster_mix} and~\ref{fig:Sk_cluster_poly} we include the corresponding data
for the static structure factors of the clusters' CMs, $S^\text{CM}_{\alpha\beta}(k)$.
The maxima of the different structure factors take moderate values,  $S^\text{CM}_{\alpha\alpha}(k) \sim 2$
for correlations between same species and $S^\text{CM}_{\alpha\beta}(k) \sim 1$ for distinct species.
No signature of Bragg peaks, which would be present in cluster crystals, is found, 
confirming the amorphous character of the arrangement of the clusters' CMs.
The significant anticorrelation, $S^\text{CM}_{\alpha\beta}(k) \sim -0.5$, of distinct species  in the low-$k$ regime
suggests a certain segregation between clusters of big and small particles. 

In summary, from the former analysis we conclude that the structure of the clusters' CMs below the clustering
transition is essentially amorphous, at least down to the lowest simulated
temperature, but differences are visible between the two investigated models (binary mixture and polydisperse system).
\ We will study the dynamic character
of these amorphous cluster phases (whether they consist of
fluids or glasses of clusters) in the next subsection.

\subsection{Dynamics}

We now turn our attention to the dynamics of the two investigated models. In doing so,
we will better characterize the nature of the cluster phases
identified in the previous section, and we will highlight the differences and
similarities with respect to the dynamics of ultrasoft particles in
cluster crystals~\cite{moreno_diffusion_2007,coslovich_hopping_2011}.

\begin{figure}[]
  \includegraphics[width=\onefig]{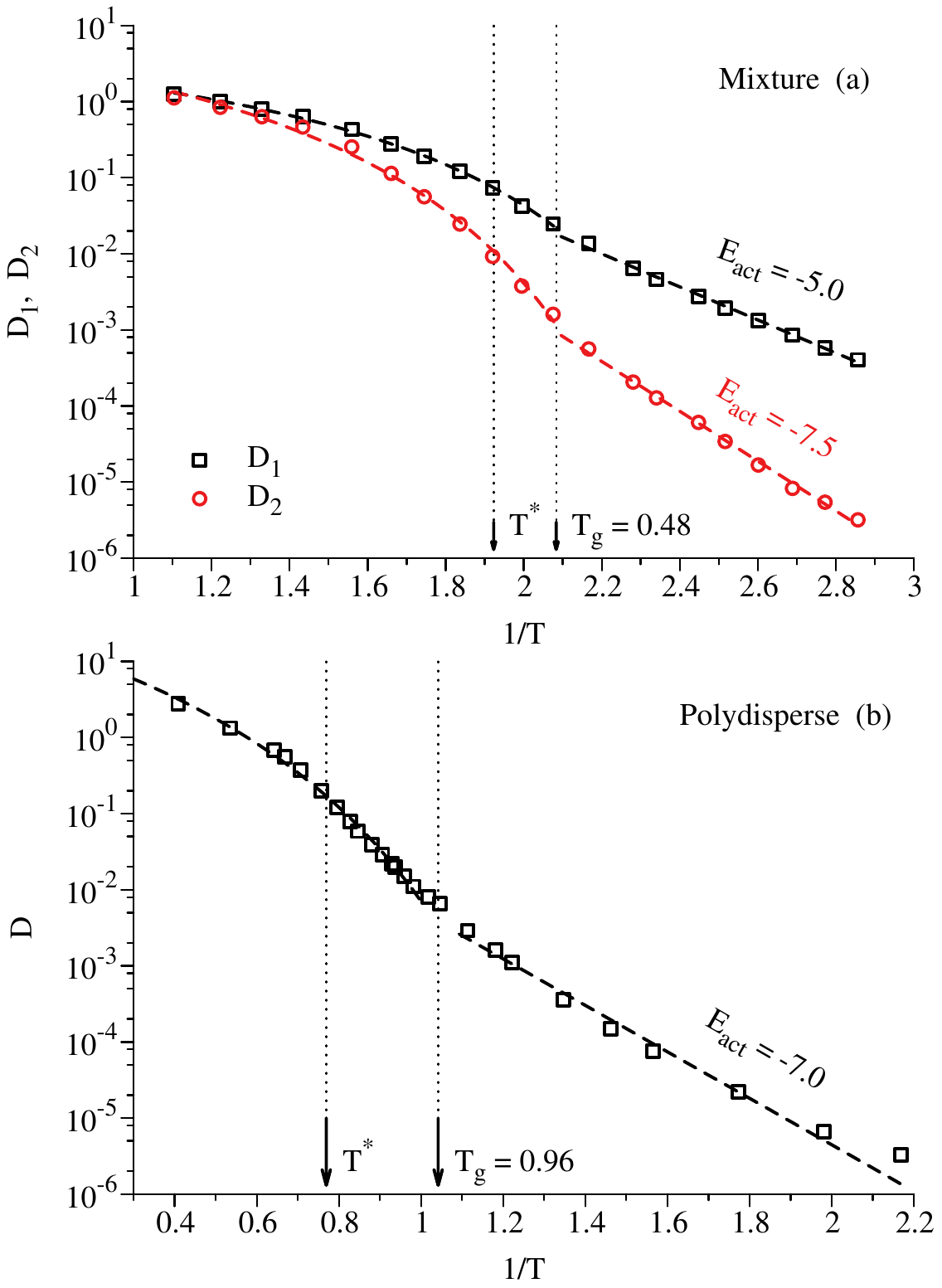}
  \caption{\label{fig:arrhenius} Arrhenius plot of the diffusion
    coefficients (symbols). (a): $D_1$ and $D_2$ for the binary mixture. (b): total
    diffusion coefficient $D$ for the polydisperse model. The vertical
    dotted lines indicate the location of the clustering transition
    ($T^*$) and cluster glass transition ($T_g$). Dashed lines are fits to an VFT law (for $T > T_{\rm g}$) and to an
    Arrhenius law (for $T < T_{\rm g}$, activation energies are indicated). }
\end{figure}

\begin{figure}[]
  \includegraphics[width=\onefig]{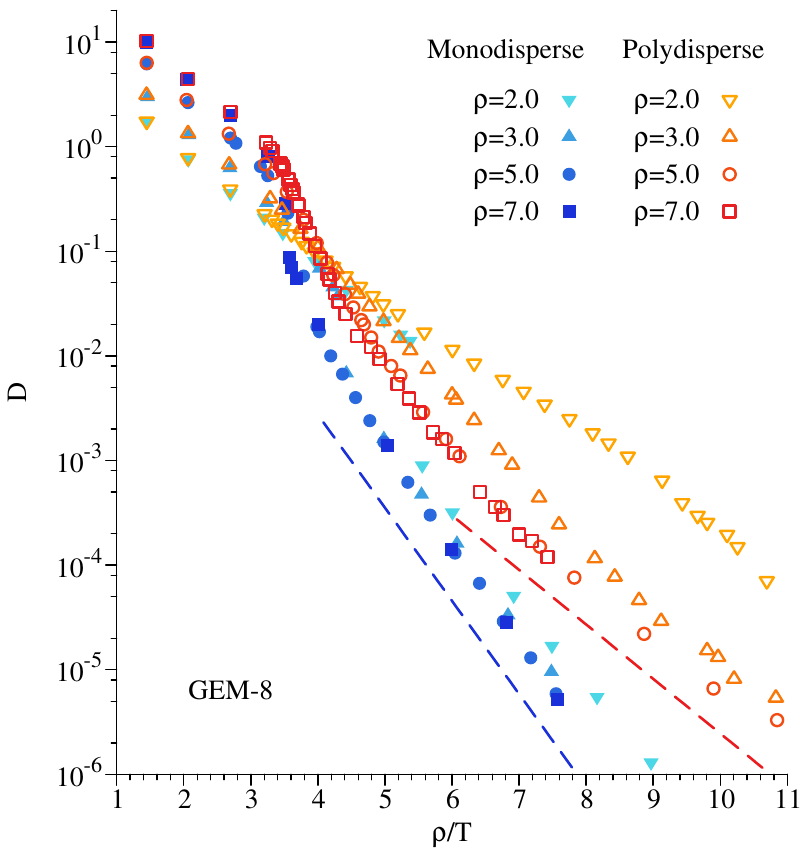}
  \caption{\label{fig:rhoT-scaling} Diffusivities of the GEM-8 model vs. $\rho/T$. Filled symbols are data for 
 the monodisperse system \cite{moreno_diffusion_2007}. Empty symbols are data for the polydisperse
  system of this work. The thick dashed lines indicate Arrhenius-like behavior in the cluster crystal and cluster glass of the monodisperse
  and polydisperse systems respectively.}
\end{figure}

Let us start with the analysis of the temperature dependences of the
diffusion coefficients. These have been extracted  from the long time limit of 
$\langle R^2_\alpha(t) \rangle /6t$,
where the partial mean square displacements are defined as:
$$\langle R^2_\alpha(t) \rangle = \big\langle \frac{1}{N_\alpha} \sum_{i=1}^{N_\alpha}|\vec{r}_i(t) -
\vec{r}_i(0)|^2 \big\rangle .$$ 
The index $i$ runs over all particles of species $\alpha$. In Fig.~\ref{fig:arrhenius} we show the
species-dependent diffusion coefficients $D_\alpha$ of the binary
mixture ($\rho = 4.0$, panel (a)) and the total diffusion coefficient $D$ of the
polydisperse model ($\rho = 5.0$, panel (b)). The very high temperature regime is omitted for clarity and will be shown in Fig.~\ref{fig:mcmd}. An Arrhenius representation is used
to highlight the development of slow dynamics upon decreasing the
temperature. A first portion of the data, covering temperatures for which the diffusion coefficients are between $\approx 1$ and  $\approx 10^{-2}$, can be reasonably well described
by a Vogel-Fulcher-Tamman (VFT) law  
$$D_\alpha(T) \sim
\exp{\left[-A^{(\alpha)}/(T-T_0^{(\alpha)})\right]}\,\, .$$ The fitted values of the
strength parameters $A^{(\alpha)}$ depend on both species and models
under consideration. We observe that the clustering temperature $T^*$
lies within this first portion of data. Thus no dynamic signature of
the thermodynamically defined clustering transition can be evidenced from this representation of the data (see also Fig.~\ref{fig:mcmd}). At lower temperatures the diffusion coefficients undergo a crossover to a milder temperature dependence, which can be
well described by an Arrhenius law $$D_\alpha(T) \sim  \exp{\left[-E^{(\alpha)}_{\rm act}/T\right]}\,\, .$$ The activation energies are
species-dependent and tend to be higher the bigger the particles.
We interpret this ``fragile-to-strong'' crossover (i.e., from VFT to Arrhenius temperature dependence) as a signature of a change
in the microscopic transport mechanisms, from the high-$T$ collective dynamics to activated, single-particle hopping taking place in a nearly frozen cluster structure. We remark that this crossover takes place below $T^*$, in a regime where clustering is nearly complete and the fraction of isolated particles $P_1$ is typically lower than 15--20\%.

\begin{figure*}[]
  \includegraphics[width=\twofig]{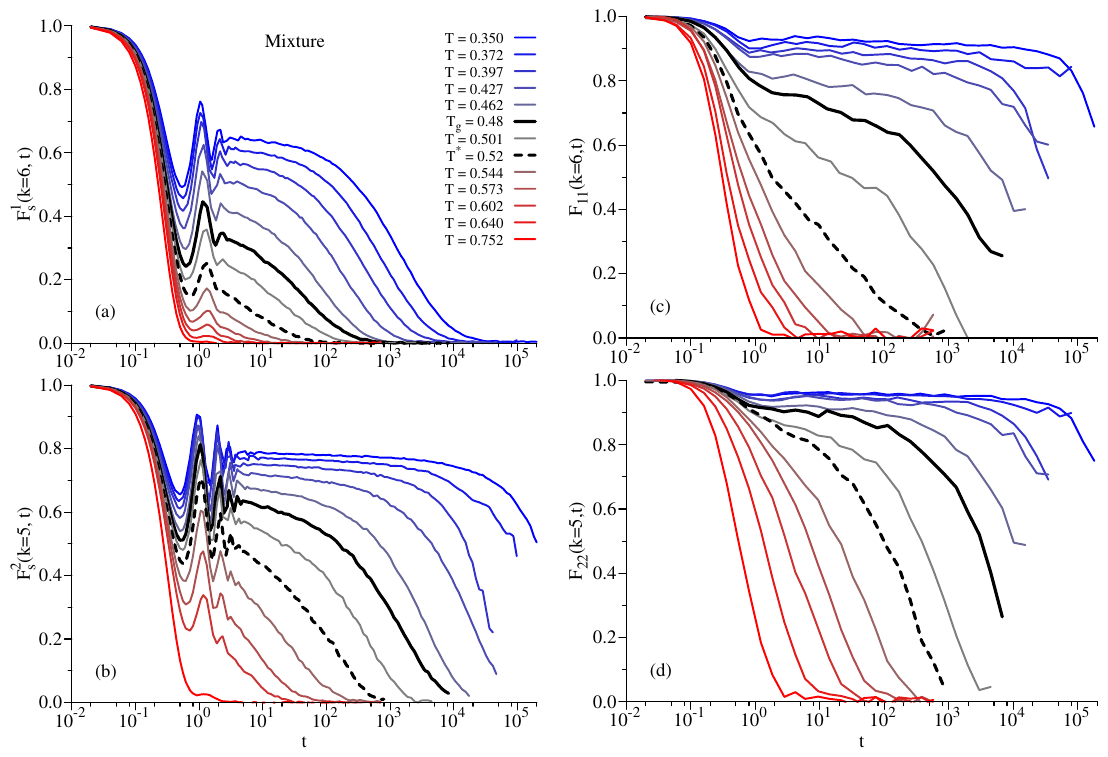}
  \caption{\label{fig:fkt_mix} Intermediate scattering
    functions for the binary mixture evaluated at various
    temperatures (see legend): (a) $F_s^1(k=6,t)$, (b) $F_s^2(k=5,t)$, (c) $F_{11}(k=6,t)$ , and (d) $F_{22}(k=5,t)$.
 The clustering and cluster glass
    transitions are highlighted with bold dashed and bold continuous
    lines, respectively.}
\end{figure*}

Figure~\ref{fig:rhoT-scaling} shows the diffusivities of the polydisperse system versus the variable $\rho/T$ at the different investigated densities. At low temperatures, data for $\rho \ge 5.0$ collapse to a common Arrhenius law, i.e., with an activation energy $E_{\rm act} \propto \rho$. Low-$T$ data for $\rho = 3.0$
do not collapse though still are close and parallel to the data for $\rho \ge 5.0$. This scaling behavior of the diffusivities
in the cluster phase of polydisperse GEM-8 particles is analogous to that observed in the fcc cluster crystal phase of the
monodisperse counterpart.  The data of the fcc cluster crystal (taken from Ref.~\cite{moreno_diffusion_2007}) are included in Fig.~\ref{fig:rhoT-scaling}
for comparison.  In analogy with the observation for the cluster crystals (see discussion in Ref.~\cite{moreno_diffusion_2007}), the Arrhenius temperature 
dependence suggests that the particles perform hopping dynamics on the sites
of an almost frozen matrix of clusters. Though the same qualitative $\rho/T$-scaling is found, quantitative differences are observed. In particular the
activation free energy in amorphous cluster phases is lower than in the cluster crystal.  This can be tentatively assigned to structural disorder leading to a larger number of available pathways (``entropic'' contribution) or to a smoother energy landscape probed by the single particles in the amorphous cluster phase, in contrast to the ordered structure of local minima separated
by high barriers in the cluster crystal~\cite{likos_cluster-forming_2008}.

The $\rho/T$-scaling of the polydisperse system breaks down at sufficiently high temperatures, i.e., above the fragile-to-strong crossover. This suggests a more complex, cooperative transport mechanism above this crossover, in analogy with the mildly supercooled regime of glass-forming liquids. We will show below that indeed the systems investigated here exhibit, in this regime, characteristic dynamic features of that scenario.
Finally, it is worth mentioning that the data for the polydisperse system and its monodisperse counterpart show a perfect overlap in the high-temperature  fluid phase, i.e, above the crystallization and clustering transition for the monodisperse and polydisperse
system respectively.  Thus, the effects of the size dispersity have no relevance for the dynamics at such high temperatures.
Consistently with the observations for the potential energy and specific heat   (Figs.~\ref{fig:cv_mix} and~\ref{fig:cv_poly}) the diffusivities
of the polydisperse system display a smooth variation around $T^*$, in contrast to those of the monodisperse system around $T_m$.
All the previous qualitative observations are also found for the $\rho/T$-scaling (obeyed for $\rho \ge 3.0$) of the diffusivities in the binary mixture (not shown).

\begin{figure}[t]
  \includegraphics[width=\onefig]{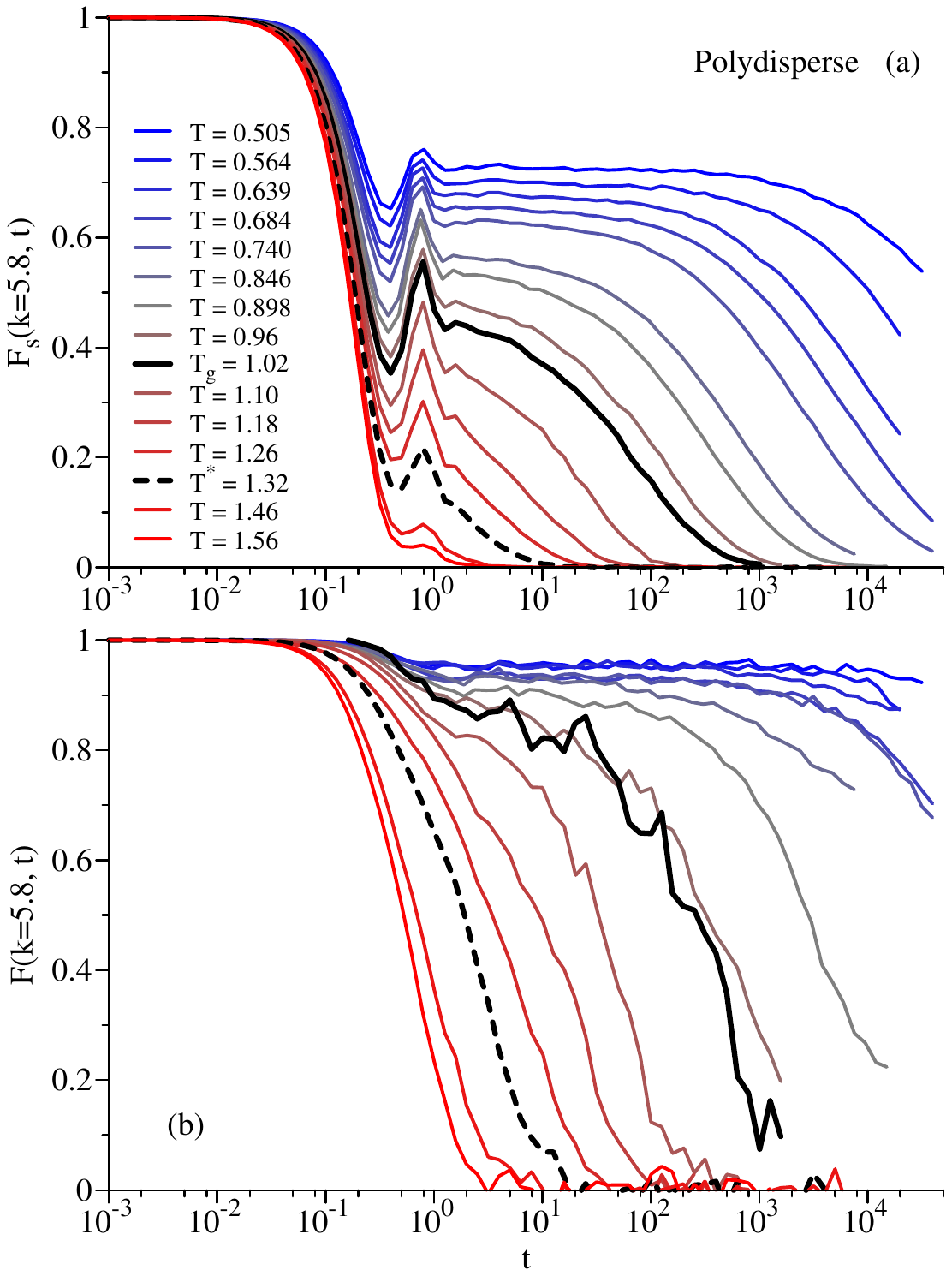}
\caption{\label{fig:fkt_poly} As fig~\ref{fig:fkt_mix} for the polydisperse model: (a) $F_s(k=5.8,t)$ and (b) $F(k=5.8,t)$. }
\end{figure}

To characterize the dynamics in more detail, we study the temperature
evolution of both incoherent and coherent intermediate scattering
functions (Figs.~\ref{fig:fkt_mix}, and~\ref{fig:fkt_poly}). In the case of the binary mixture, we
report separately the scattering functions for small and large
particles, whereas an average over all particles is performed for the polydisperse model. The selected wave vectors reflect
the positions of the respective first peaks in the static structure factors.
For the binary mixture the values are $k^*=6.0$ (for $\alpha=1$) and $k^*=5.0$ (for $\alpha=2$),
whereas  for the polydisperse model  $k^*=5.8$. 

We first discuss the incoherent scattering functions
$F_s(k,t)$.  These are calculated as 
\begin{equation}
F^{\alpha}_s(k,t) = \frac{1}{N_{\alpha}}\sum_{j=1}^{N_\alpha} \exp[i \vec{k}\cdot (\vec{r}_j(t)  -\vec{r}_j(0) )] ,
\end{equation}
where the sum is performed over the positions,  $\vec{r}_j$, of the $N_\alpha$ particles of the species $\alpha$.
At high temperature the correlation functions decay
rapidly to zero in a simple exponential fashion. Around and below
$T^*$, marked oscillations appear at short times $t \sim 1$, which can be
attributed to the vibrations of individual particles within the
clusters. Similar features have been observed in the cluster
crystal phases of similar ultrasoft particles~\cite{coslovich_hopping_2011}, and are known to be associated
to single-particle vibrational modes~\cite{Neuhaus_Likos_2011}. As the temperature is further
decreased, the damping of these oscillations becomes weaker and the
amplitude gets larger, consistently with an increased stability of the
clusters. The oscillations are also visible in the
polydisperse model (see Fig.~\ref{fig:fkt_poly}a), though they are less pronounced due to the average 
over all particles' sizes.

The appearance of oscillations in $F_s(k,t)$ is accompanied by
evident signatures of glassy dynamics. A plateau in $F_s(k,t)$
develops at intermediate times and apparently grows in a continuous fashion from zero to
finite values. Such an increase of the plateau height indicates a
dynamical slowing down of the continuous type. Interestingly, this feature resembles  the type-A
transition scenario predicted by the Mode Coupling Theory for certain
classes of glassy systems~\cite{gotze_recent_1999,Gotze_2009}. It should be noted that the emergence
of the plateau occurs clearly above $T^*$, i.e., the onset of 
slow dynamics already takes place prior to the thermodynamically defined clustering transition,
both for the binary mixture and for the polydisperse system. This is consistent with the smoothness of the transition
(see Figs.~\ref{fig:cv_mix} and ~\ref{fig:cv_poly}). Concomitant with the observed decrease of the diffusivities observed in Fig.~\ref{fig:arrhenius},
the plateau extends over longer time scales  and is followed by a slow decay
with increasing relaxation time as temperature decreases.

\begin{figure}[]
  \includegraphics[width=\onefig]{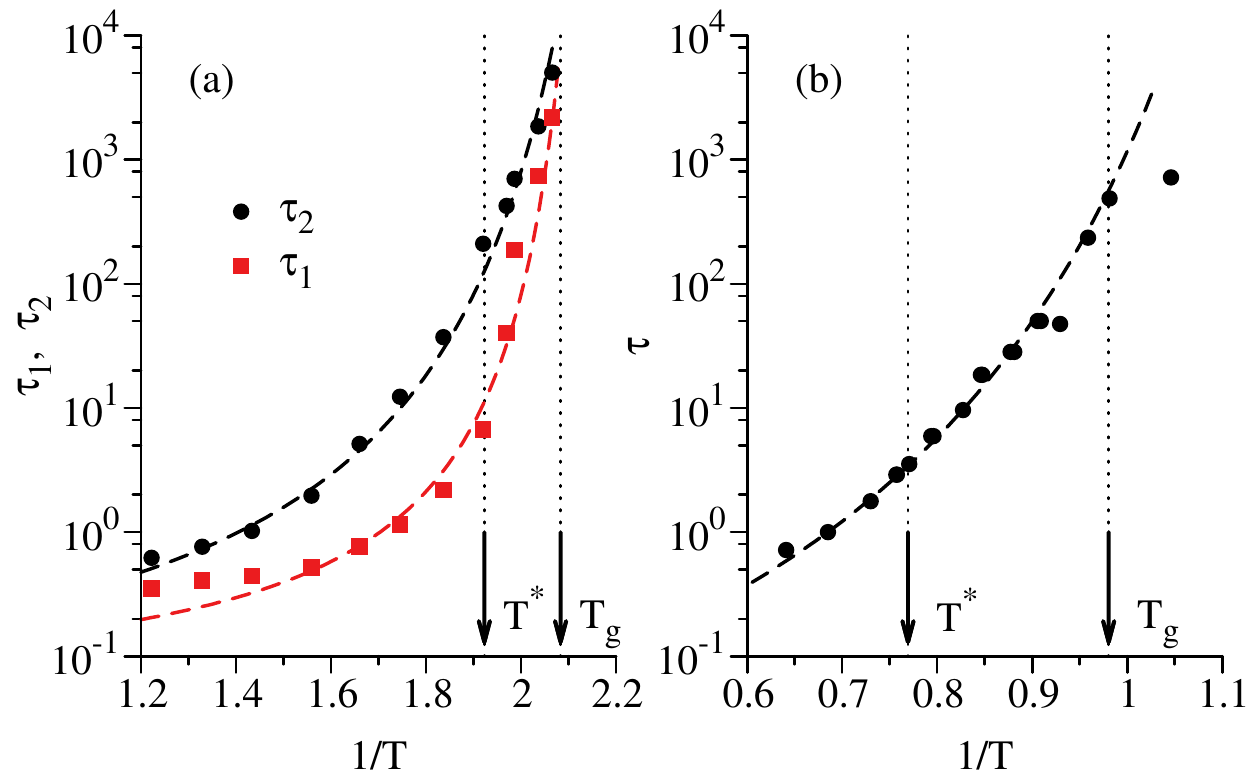}
  \caption{\label{fig:tau} Structural relaxation times (a) $\tau_1$ and $\tau_2$ for the binary mixture and (b) $\tau$ for the polydisperse model as a function of $1/T$. The dashed lines represents VFT fits.}
\end{figure}

The coherent scattering functions are calculated as 
\begin{equation}\label{eqn:fkt}
F_{\alpha\beta}(k,t) = 
\frac{\langle \rho_\alpha(k,t)\rho_\beta(-k,0)\rangle}{\langle \rho_\alpha(k,0)\rho_\beta(-k,0) \rangle} ,
\end{equation}
where $\rho_\alpha(k,t)=\sum_{j=1}^{N_\alpha} \exp{[i \vec{k}\cdot\vec{r}_j(t)]}$. 
Inspection of the coherent functions $F(k^*,t)$ of the binary mixture (Fig.~\ref{fig:fkt_mix}c-d) and the polydisperse model (Fig.~\ref{fig:fkt_poly}b) reveals similar features of glassy dynamics.
Since coherent functions probe collective correlations, the data of Figs.~\ref{fig:fkt_mix}c-d and~\ref{fig:fkt_poly}b reflect a progressive dynamic arrest of the cluster structure. At very low temperature, $F(k^*,t)$ does not relax to zero within our observation times and the cluster structure is effectively frozen. The highest temperature at which $F(k^*,t)$ does not completely relax to zero over the available time scale defines the \textit{cluster glass} transition and will be denoted as $T_g$. We find $T_g\approx 0.48$ and $T_g\approx 1.02$ for the binary mixture and the polydisperse model, respectively~\footnote{Even though $T_g$ is in striking agreement with the  fragile-to-strong crossover temperature, we emphasize that the cluster glass transition is a manifestation of the dynamic arrest of the clusters' centers-of-mass and should thus be defined in terms of collective properties. Therefore we prefer not to identify $T_g$ with the fragile-to-strong crossover temperature.}. 

\begin{figure*}[!t]
  \includegraphics[width=0.85\twofig]{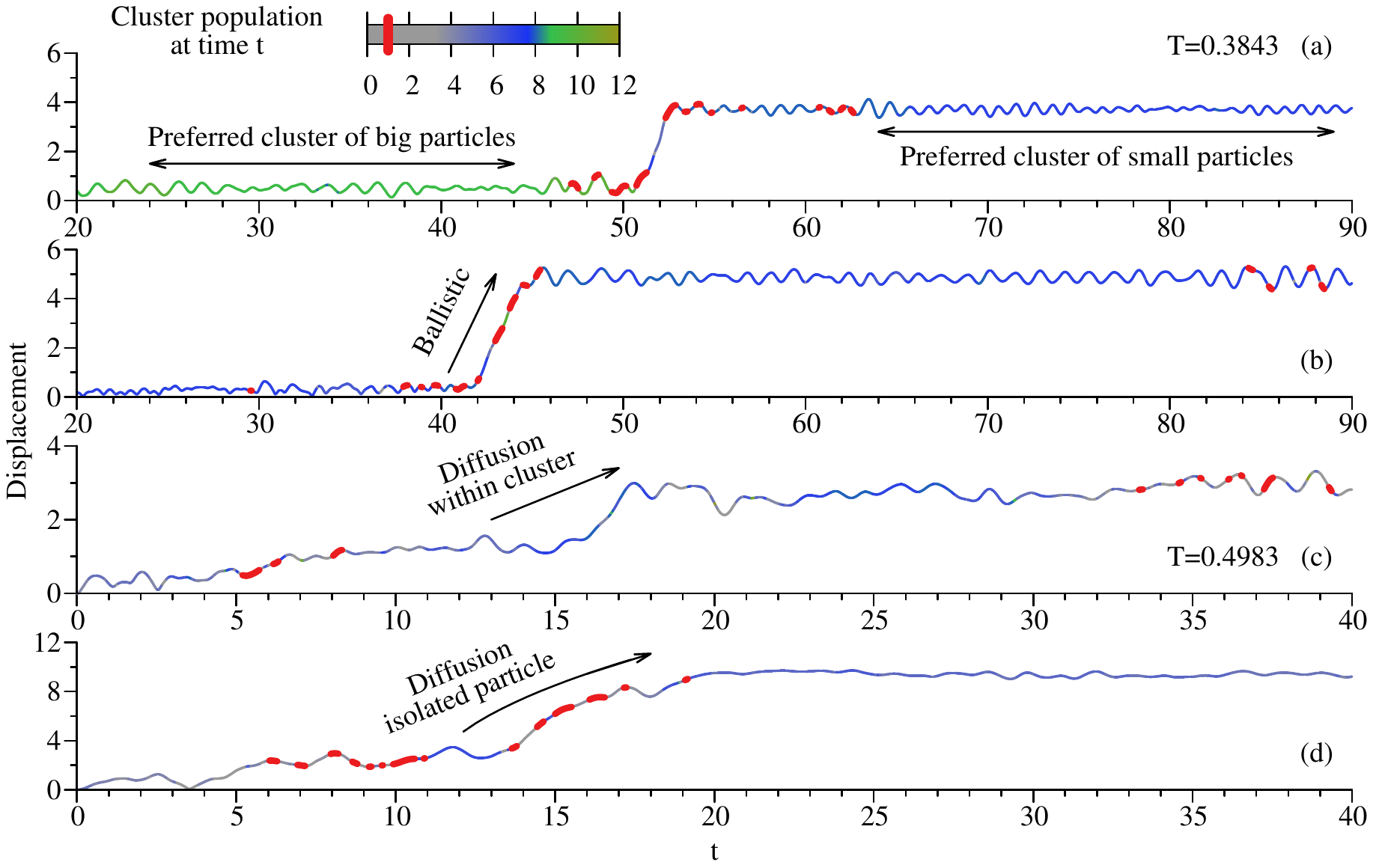}
\caption{\label{fig:traj} Typical displacements $|\vec{r}_i(t)-\vec{r}_i(t_0)|$ of selected small particles in the binary mixture at $T=0.3843$ (panels (a) and (b)) and $T=0.4983$  (panels (c) and (d)). As indicated in the legend, the color of the line indicates the population of the cluster to which the particle belongs at time $t$. Portions of the trajectory during which the particle is isolated are highlighted with a thick red line.}
\end{figure*}

\begin{figure*}[]
  \includegraphics[width=\twofig]{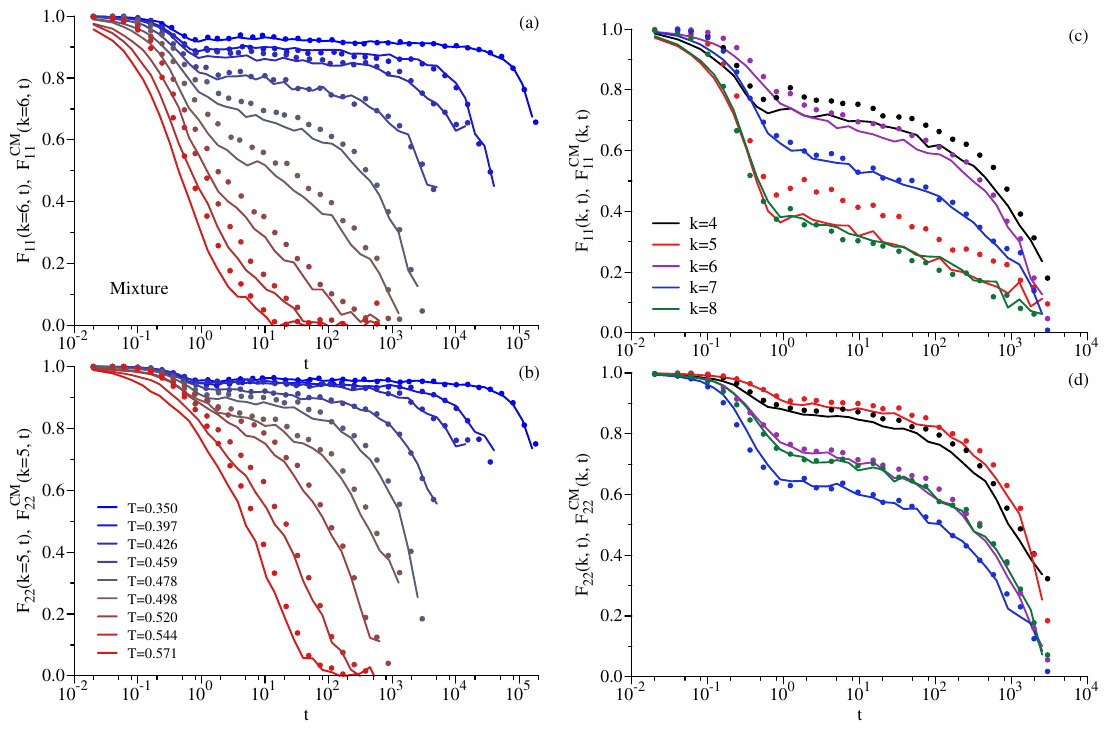}
\caption{\label{fig:fkt_cluster_all_mixture} 
Symbols: coherent scattering functions $F_{\alpha\beta}^\text{CM}(k^*,t)$  evaluated for the
  clusters' centers of mass  in the binary mixture. The respective data for the coherent scattering functions $F_{\alpha\beta}(k,t)$, calculated on a particle basis,
  are included as solid lines. Panels (a) and (b) show data at different temperatures (see legend) and fixed wave vector. The selected wave vectors are $k^* = 6$ for 1-1 correlations (a)
  and $k^* = 5$ for 2-2 correlations (b).  Panels (c) and (d) show data at fixed temperature $T=T_g=0.48$ and different wave vectors (see legend) for 1-1 correlations (c)
  and 2-2 correlations (d).}
\end{figure*}

The slowing down of $F(k^*,t)$ above $T_g$ follows the two-step scenario typically observed in supercooled liquids. This allows one to define, as usual, the corresponding structural relaxation times $\tau$ from the condition $F(k^*,\tau)=1/e$. Figure~\ref{fig:tau} shows the temperature variation of the species-dependent relaxation times $\tau_1$ and $\tau_2$ for the binary mixture and the total relaxation time $\tau$ for the polydisperse model. The data have been tentatively fitted using the VFT function, yielding values of $T_0$ in the range 0.42--0.45 for the binary mixture and 0.65--0.7 for the polydisperse model. These temperatures can be taken as lower bounds for the cluster glass transition, corresponding to infinitely slow cooling rates. Furthermore, the separation between $T^*$ and $T_0$ is very pronounced in the polydisperse model, which indicates that the cluster glass formed by the latter is ``stronger'', in the Angell's terminology~\cite{angell_formation_1995-1}, than the one formed by the binary mixture. All this clearly shows that the clustering and cluster glass transitions are distinct and different in nature: the former is an equilibrium crossover, the latter an ergodicity breaking transition that would occur at different temperatures depending on observation times.

Let us summarize the scenario emerging from our simulations. Consistent with the structural features discussed in subsection~\ref{sec:results}A, the clustering transition is associated to a crossover from a fluid of mostly dissociated particles to a \textit{fluid of clusters}. The dispersity of the clusters evidenced in~\ref{sec:results}A frustrates crystallization of the clusters CMs and allows quenching the fluid to lower temperatures. In the temperature range $T_g \alt T \alt T^*$ the underlying transport processes are rather complex: by visual inspection of animated particles' trajectories, we observed diffusive motion of clusters as a whole, branching of clusters as well as random walks of isolated particles. This results in non-Arrhenius temperature dependence of the transport coefficients and a non-trivial behavior of the intermediate scattering functions.
Below $T_g$ the structure of the clusters' centers-of-mass becomes practically arrested on the time scale of our simulation (see below) and the system enters in the cluster glass
regime. Having noted this, the incoherent scattering functions still decay to zero in the time scale of the simulation at some temperatures $T < T_g$ (see Figs.~\ref{fig:fkt_mix}a-b
and~\ref{fig:fkt_poly}a). In other words, single-particle dynamics is still effectively ergodic in the cluster glass and is driven by hopping between the effectively arrested
clusters. This hopping dynamics is manifested by the Arrhenius-like behavior observed in the diffusivities (see Fig.~\ref{fig:arrhenius}). Therefore, as in cluster
crystals~\cite{moreno_diffusion_2007}, single-particle and collective dynamics in the cluster glass are strongly decoupled. Representative particles' displacements illustrating the
difference in transport below and above $T_g$ are depicted in Fig.~\ref{fig:traj}a-b and Fig.~\ref{fig:traj}c-d, respectively. Panels (a) and (b) show typical single-particle
hopping processes at low $T$, involving jumps over a few neighbor distances. Note that, during the jumps, the motion of the particle is nearly ballistic, 
as in cluster crystals~\cite{coslovich_hopping_2011}, and may involve passing through different clusters. Panels (c) and (d) illustrate the more complex motions occurring above $T_g$, which involve both slow diffusion of particles residing \textit{within} clusters (panel (c)) and diffusion of isolated particles (panel (d)).

To conclude this section, we provide now explicit evidence that the cluster glass transition corresponds to the loss of ergodicity in the degrees of freedom associated to the clusters' centers of mass. At each  time $t$ we identify the positions $\vec{R}_j(t)$ of the clusters' centers of mass, where $1 \le j \le N_{cl}(t)$ and $N_{cl}(t)$ is the total number of clusters in the system at time $t$. We then calculate the coherent scattering functions of the clusters' CMs as
\begin{equation}\label{eqn:fkt_cm}
F^\text{CM}_{\alpha\beta}(k,t) = 
\frac{\langle \rho_\alpha^\text{CM}(k,t)\rho_\beta^\text{CM}(-k,0) \rangle}{\langle \rho_\alpha^\text{CM}(k,0)\rho_\beta^\text{CM}(-k,0) \rangle} ,
\end{equation}
where $\rho^\text{CM}_\alpha(k,t)=\sum_{j=1}^{N^{\alpha}_{cl}(t)} \exp{[i
  \vec{k}\cdot\vec{R}^{\alpha}_j(t)]}$ and the sum is done over the $N_{cl}^{\alpha}(t)$ clusters
of the species $\alpha$. The chemical species $\alpha$ of
a given cluster is again defined as the majority species in the cluster (see above). Single
clusters (i.e., isolated particles)  are excluded from this analysis. It is
worth mentioning that the averages in Eq.~\eqref{eqn:fkt_cm} are
performed in a grand canonical ensemble, since the number of clusters
in the sample fluctuates in time. A comparison between
$F_{\alpha\beta}(k,t)$ and $F_{\alpha\beta}^\text{CM}(k,t)$ of the binary mixture is shown
in Fig.~\ref{fig:fkt_cluster_all_mixture}(a-b), for wave vectors at the
first peak in the corresponding static structure factors, and at different temperatures. Both data sets follow closely each other.
When looking at the former scattering functions at some fixed, low temperature, but changing the wave-vector
(Fig.~\ref{fig:fkt_cluster_all_mixture}(c-d)) the agreement remains overall
good, especially between big particles and big clusters (2-2 correlations). With this, 
we conclude that the collective slowing down of the fluid is indeed driven by
the progressive arrest of the clusters' structure.

\subsection{Dependence on thermal history and microscopic dynamics}\label{sec:qrate}

\begin{figure}[t]
  \includegraphics[width=\onefig]{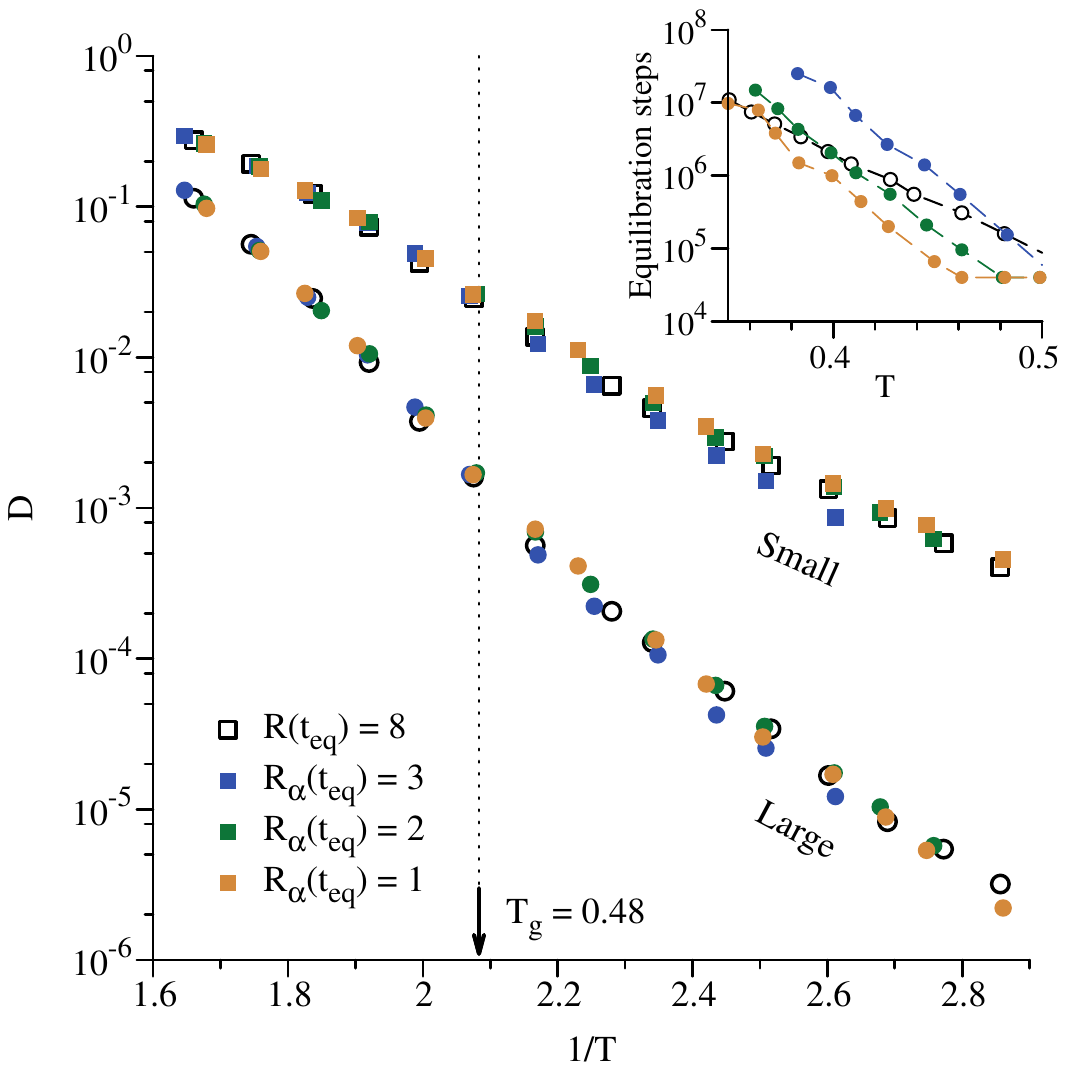}
\caption{\label{fig:arrhenius_qrate} $T$-dependence of the diffusion
  coefficients $D_1$ and $D_2$ in the binary mixture for different equilibration criteria. The different
  data sets correspond to different values of the total RMSD (open symbols) and partial RMSD (full symbols) targeted during
  the equilibration run (see legend). Inset:
  equilibration time $t_{eq}$ as a function of $T$ for the different
  equilibration criteria.}
\end{figure}

\begin{figure}[!t]
  \includegraphics[width=\onefig]{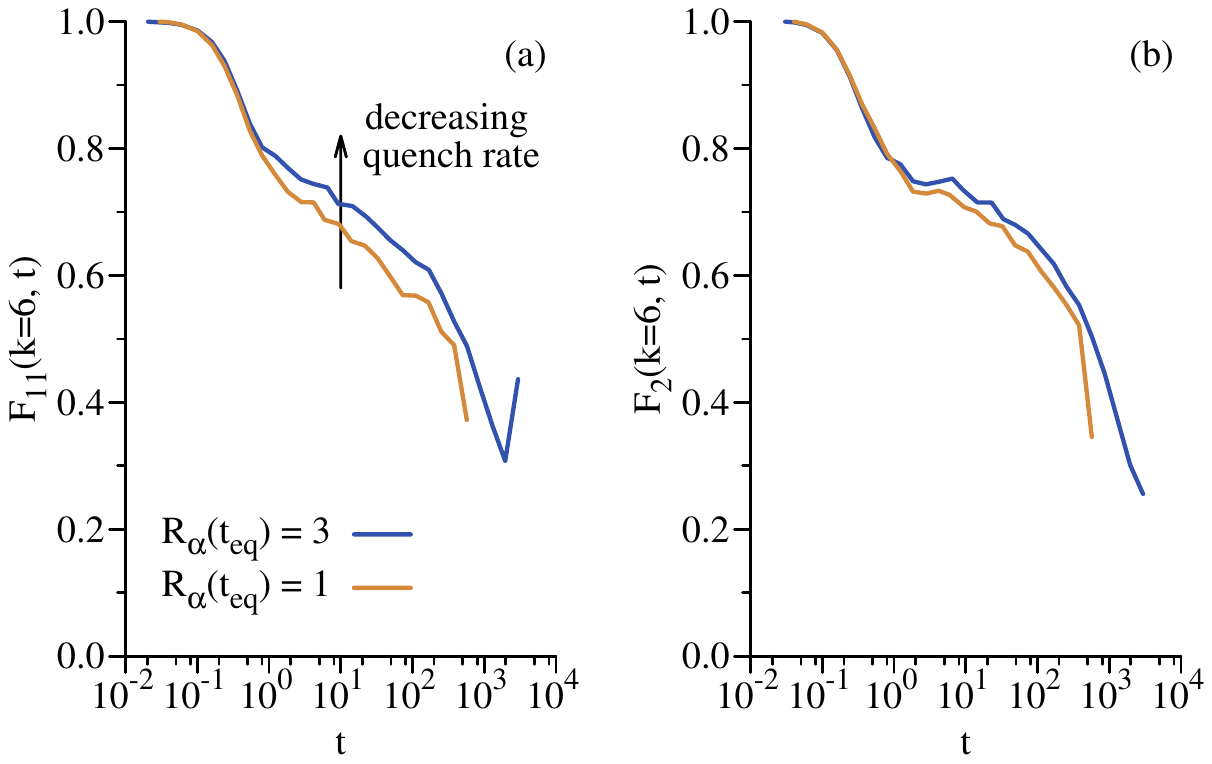}
  \caption{\label{fig:fkt_qrate} Dependence of the coherent intermediate scattering functions (a) $F_{1}(k=6,t)$ and (b) $F_{2}(k=6,t)$ on the equilibration criterion in the binary mixture at $T=T_g=0.48$. Colors are the same as in Fig.~\ref{fig:arrhenius_qrate}.}
\end{figure}

To corroborate our interpretation of $T_g$ as a glass transition occurring at the
clusters' level, we now study how the clusters' structure and dynamics depend on
the quench history
obtained from three different thermal histories for the binary mixture. The
difference between the data sets considered in this section lies in the time $t_\text{eq}$ allowed
for equilibration: the equilibration runs are such that the partial RMSD $R_\alpha(t_\text{eq})$, evaluated at the end of the run, is always larger than one, two, and three, respectively. Note that this equilibration criterion is slightly different from the one employed in the rest of the work (indicated in Fig.~\ref{fig:arrhenius_qrate} for comparison), which is based on the \textit{total} RMSD of particles. Production runs are typically 4 times longer than the corresponding equilibration runs~\footnote{In the case of the polydisperse model, we could not observe any marked dependence of static and dynamic properties on equilibration times, although we did not perform a systematic analysis as for the binary mixture.}. In Fig.~\ref{fig:arrhenius_qrate} we analyze the influence of the different quench rates on the $T$-dependence of the diffusion coefficients. We remark that, since the equilibration times are adjusted at each thermodynamic state to obtain the target RMSD, a measure of the average quench rate is not appropriate. In fact, as shown in the insets of Fig.~\ref{fig:arrhenius_qrate}, the equilibration times increase exponentially with decreasing $T$. 
Within statistics the three data sets of Fig.~\ref{fig:arrhenius_qrate} show a perfect
overlap for $T > T_g$. However, small but discernible differences 
are visible below $T_g$ (error bars are of the order of the symbol size). Thus, longer equilibration runs lead to slightly lower diffusion coefficients. A similar and more pronounced effect is visible in the relaxation of the coherent intermediate scattering functions, which are shown in Fig.~\ref{fig:fkt_qrate} at $T=0.48=T_g$ for two different quench rates. Not only $\tau$ increases with increasing $R_\alpha(t_\text{eq})$, but also the extent to which $F_\alpha(k,t)$ has relaxed during the simulation increases with increasing $R_\alpha(t_\text{eq})$, hence with the total length of the production run. This, in turn, shifts the cluster glass transition to lower temperatures. 

Whereas at the superficial level these observations resemble the typical aging process in structural glasses, the microscopic origin of the quench rate dependence of the diffusion coefficients is not trivial. In fact, at least two factors may affect the single particle dynamics below the cluster glass transition. First, the structure of the clusters' CM relaxes faster upon decreasing equilibration times. The single particle dynamics, in turn, is affected by the relaxation of the cluster structure, since some relaxation channels (e.g., diffusion of a cluster as a whole), which would be suppressed upon longer annealing, remain active and increase the diffusivity.

\begin{figure}[!t]
  \includegraphics[width=\onefig]{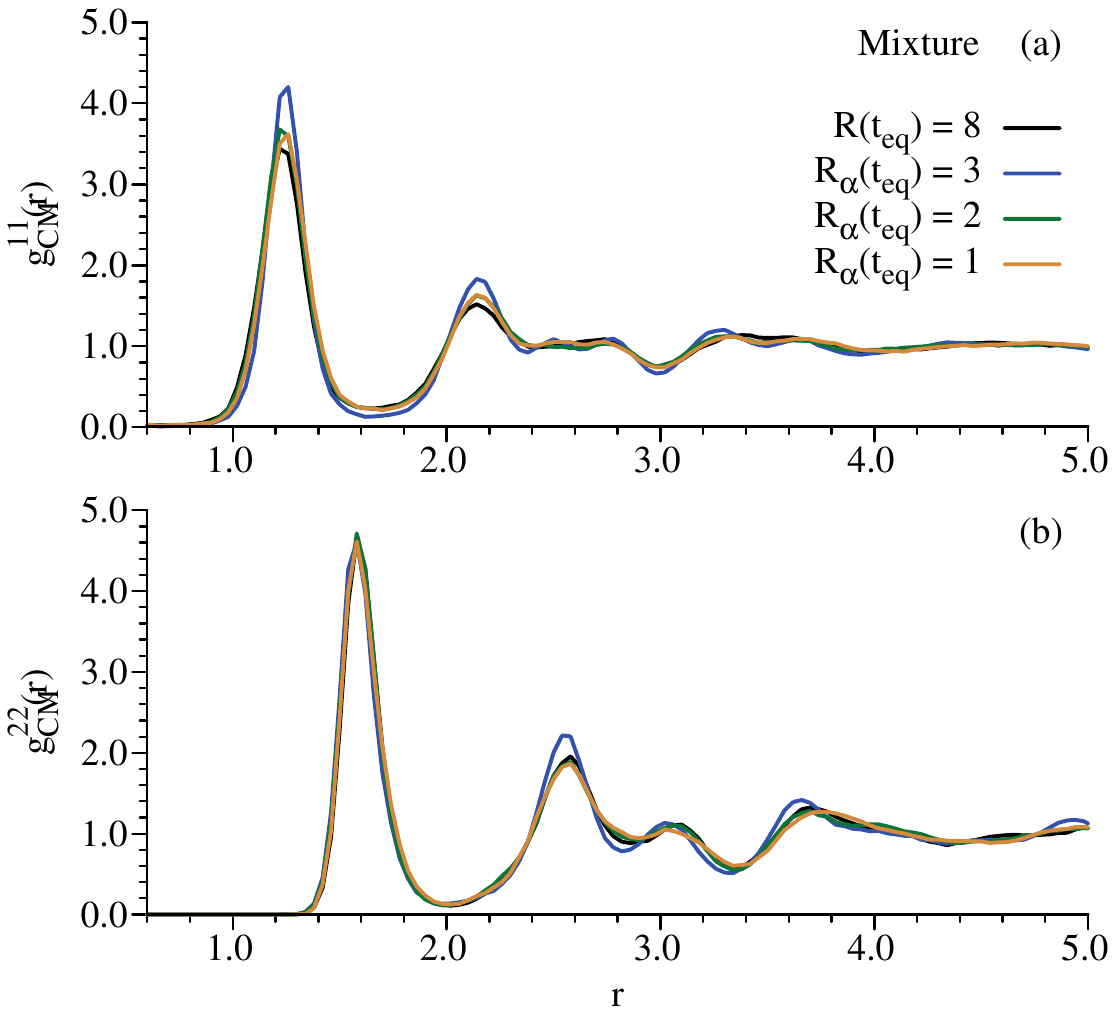}
  \caption{\label{fig:gr_qrate} Dependence of the radial
  distribution functions (a) $g^\text{CM}_{11}(r)$ and (b) $g^\text{CM}_{22}(r)$ on the equilibration
  criterion in the binary mixture at $T=0.3843<T_g$. Colors are the same as in Fig.~\ref{fig:arrhenius_qrate}.}
\end{figure}

One may also speculate that a further contribution to the single particles' mobility arises from differences in the \textit{topology} of the cluster glass, which
determines in turn the available pathways for particles' hopping.  To
assess whether this is a possible scenario, we analyze the radial
distribution functions $g^\text{CM}_{11}(R)$ of the
clusters' CM in the binary mixture as a function of the equilibration criterion
(see Fig.~\ref{fig:gr_qrate}). We see that the cluster structure of the
most slowly annealed sample differs appreciably from the others. Though the positions of the peaks
are unchanged, their amplitude is increased, indicating a
more pronounced ordering. These somewhat large differences can be explained by the
fact that, as evidenced in the inset of Fig.~\ref{fig:arrhenius_qrate},
the average quench rate does not change linearly with the value of $R_\alpha(t_\text{eq})$.~\footnote{One might argue that differences in the cluster structure for the longest annealing originate from partial crystallization. However, neither the analysis of thermodynamical properties ($U$, $P$, $C_V$) nor visual inspection of the clusters'
structure revealed evident signatures of crystallization. Irrespective
of the fate of the slowly annealed sample and its possible
crystallization, data in Fig.~\ref{fig:gr_qrate} evidence
that the clusters' structure is more ordered for the longer annealed samples.} From the comparison of cluster crystals and cluster glasses (see Fig.~\ref{fig:rhoT-scaling}), we see that hopping in a more disordered matrix decreases the activation barrier for diffusion, and hence increases the particles' mobility. The actual dependence of $D$ on the quench rate shown in Fig.~\ref{fig:arrhenius_qrate} suggests therefore that a similar effect might be at play in this model. More systematic investigations
should be performed to assess how the specific structure of the disordered matrix affects the
transport properties at the single particle level.

\begin{figure}[]
  \includegraphics[width=\onefig]{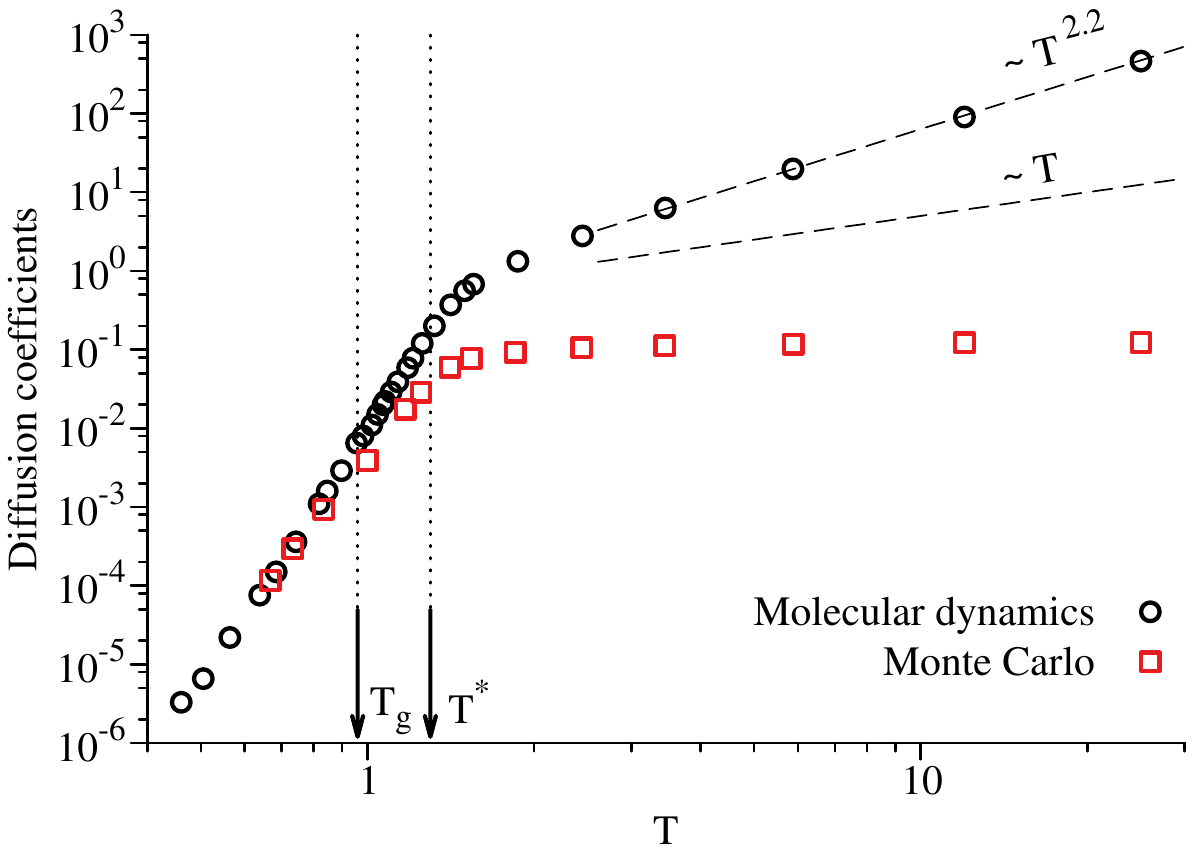}
  \caption{\label{fig:mcmd} Comparison of the temperature
  dependence of the diffusion coefficients obtained from Newtonian
  (circles) and Monte Carlo (squares) dynamics in the polydisperse system. The diffusion
  coefficients of the MC data set have been multiplied by a factor 75.
 The vertical dotted lines indicate the clustering transition $T^*$ and glass transition $T_{\rm g}$.
 The dashed lines indicate power-law  and linear behavior, and are included for comparison with the simulation data.}
\end{figure}

Finally, we briefly analyze the role of the microscopic dynamics on
the transport properties. Following~\cite{coslovich_hopping_2011} we
compare the diffusion coefficients obtained from Newtonian and Monte
Carlo dynamics for the polydisperse model. The results of this
analysis in the polydisperse system are shown in Fig.~\ref{fig:mcmd}. For temperatures
above the clustering transition the diffusion coefficients obtained
from the two methods differ qualitatively above $T^*$, in that the MC data
saturate at high temperature, whereas the MD data show a
monotonic increase. Such differences are expected, since at high $T$
the GEM-$n$ interactions will be negligible (random walk on a nearly flat energy landscape). 
Therefore in that limit the MC dynamics will become purely random and $T$-independent.
On the contrary, in Newtonian dynamics the onset time of the crossover from ballistic to diffusive motion in the mean square displacement increases systematically with increasing temperature, 
and so does  the diffusivity. Interestingly, in this regime, the temperature dependence of $D$ is approximately given by a power law $T^\nu$, with $\nu \approx 2.2$. This value of $\nu$ is higher than the ones expected for both ideal gas ($\nu=0.5$) and Brownian motion ($\nu=1$), and remains to be understood.

Only below $T_g$ (i.e., in the cluster glass) the data sets obtained from the
two dynamics can be collapsed reasonably well onto a unique master
curve, by appropriately rescaling the time units. With this, the microscopic dynamics does not seem
to play a relevant role in the transport properties of the cluster glass.
This result is rather different from the observation in cluster crystal phases \cite{coslovich_hopping_2011}. 
In such systems the Newtonian dynamics leads to long persistent jumps over the lattice sites.
These highly directional motions are suppressed in the MC dynamics, and the corresponding
MD and MC diffusivities in the cluster crystal exhibit a qualitatively different $T$-dependence,
i.e, they cannot be mutually rescaled by a $T$-independent time factor \cite{coslovich_hopping_2011}.
The fact that, instead, scaling of Newtonian and MC dynamics works reasonably well in the cluster glass 
suggests that the disordered structure of the clusters' centers-of-mass partly
suppresses the mentioned long jumps in Newtonian dynamics. 
A more systematic analysis is required to completely settle this issue. Work in this direction is in progress.

 \section{Conclusions}\label{sec:conclusions}

We have presented a computational investigation of the structural, thermodynamic and dynamic features
of dense fluids of ultrasoft fully-penetrable particles. 
The investigated systems are a binary mixture and a polydisperse system of particles interacting via the generalized exponential model,
for which equilibrium cluster crystal phases exist in the monodisperse case.
Because of the introduced dispersity, the systems investigated in this work form disordered cluster phases. 

We have characterized the  structure of the clusters.  The analysis reveals a microsegregation of the big and small particles,
and  a strong homo-coordination in the case of the binary mixture. 
The analysis of the thermodynamic observables does not provide yet evidence for a thermodynamic transition associated
to the clustering  transition. Instead, the clustering transition appears as a smooth crossover to a regime in which 
most of the particles are  located in clusters, isolated particles being infrequent. The structure of the clusters' centers-of-mass mirrors that of the dissociated fluid: the binary mixture effectively self-assemble into a binary mixture of clusters, whereas in the polydisperse model the clusters remain polydisperse in population and size.

The dispersity of the clusters drives a progressive dynamical arrest at the level of the clusters' centers-of-mass on approaching the ``cluster glass transition'', $T_g$. The latter is operationally defined by inspection of the coherent scattering functions, which exhibit characteristic features of glass-forming liquids above $T_g$ and do not completely relax to zero below $T_g$. The diffusivities exhibit 
a fragile-to-strong crossover upon decreasing temperature. The onset of strong (Arrhenius-like) behavior occurs 
below the clustering transition and signals a change in the transport mechanisms. Relaxation below the cluster glass transition is driven by particle hopping between the nearly arrested clusters. 
The analysis of the dependence of dynamic and structural properties on the equilibration times confirms the glassy nature of the systems at low temperatures. Finally, we have investigated the role of the microscopic dynamics in the transport properties, by comparing results from Newtonian and Monte Carlo simulations. At low temperature, the diffusivities obtained by both methods can be related reasonably well by a single scaling factor. This suggests that the microscopic dynamics might play a less significant role for cluster glasses
than for cluster crystals. In view of the recent observation of clustering in fully atomistic models of interpenetrable colloidal particles~\cite{lenz_monomer-resolved_2011}, we believe disordered cluster phases of ultrasoft particles deserve deeper and more systematic investigations.

\begin{acknowledgements} 

We acknowledge financial support from the projects MAT2007-63681 (Spain) and IT-436-07 (GV, Spain).
We acknowledge Centro de Supercomputaci\'{o}n de Catalu\~{n}a (Barcelona, Spain) and the HPC@LR Center of Competence in High-Performance Computing of Languedoc-Roussillon (France) for allocation of CPU time.
We thank A. Ikeda, G. Kahl, and C.N. Likos  for valuable discussions.

\end{acknowledgements}


%

\end{document}